\documentclass{mn2e}
\usepackage{color}

\newcommand{\clues}{{\sc CLUES~}}
\newcommand{\be}{\begin{equation}}
\newcommand{\ee}{\end{equation}}

\newcommand{\hMpc}{{\ifmmode{h^{-1}{\rm Mpc}}\else{$h^{-1}$Mpc}\fi}}
\newcommand{\hkpc}{{\ifmmode{h^{-1}{\rm kpc}}\else{$h^{-1}$kpc}\fi}}
\newcommand{\hMsun}{{\ifmmode{h^{-1}{\rm {M_{\odot}}}}\else{$h^{-1}{\rm{M_{\odot}}}$}\fi}
}
\newcommand{\Msun}{{\ifmmode{{\rm {M_{\odot}}}}\else{${\rm{M_{\odot}}}$}\fi}}

\def\lesssim{\mathrel{\hbox{\rlap{\hbox{\lower4pt\hbox{$\sim$}}}\hbox{$<$}}}}
\def\gtrsim{\mathrel{\hbox{\rlap{\hbox{\lower4pt\hbox{$\sim$}}}\hbox{$>$}}}}

\input{epsf}
\title[The grouping, merging and survival of subhaloes]
{The grouping, merging and survival of subhaloes in the simulated Local Group}

\author[J. Klimentowski et al.]
    {Jaros{\l}aw Klimentowski,$^{1}$ Ewa L. {\L}okas,$^{1}$ Alexander Knebe,$^{2}$ Stefan Gottl\"ober,$^{3}$\newauthor
    Luis A. Martinez-Vaquero,$^{2}$ Gustavo Yepes$^{2}$ and Yehuda Hoffman$^{4}$
    \\
    \\
    $^1$Nicolaus Copernicus Astronomical Center, Bartycka 18,
    00-716 Warsaw, Poland\\
    $^2$Grupo de Astrof{\'\i}sica, Departamento de Fisica Teorica, Modulo C-XI, Facultad de Ciencias,
	Universidad Autonoma de Madrid,\\ 28049 Cantoblanco, Madrid, Spain\\
    $^3$Astrophysikalisches Institut Potsdam, An der Sternwarte 16, 14482 Potsdam, Germany\\
    $^4$Racah Institute of Physics, Hebrew University, Jerusalem 91904, Israel
    }
\begin{document}

\maketitle

\begin{abstract}
We use a simulation performed within the Constrained Local UniversE Simulation (CLUES) project to study a realistic
Local Group-like object. We employ this group as a numerical laboratory for studying the evolution of the population
of its subhaloes from the point of view of the effects it may have on the origin of different types of dwarf galaxies.
We focus on the processes of tidal stripping of the satellites, their interaction, merging and grouping before infall.
The tidal stripping manifests itself in the transition between the phase of mass accretion and mass loss seen in most
subhaloes, which occurs at the moment of infall on to the host halo, and the change of the shape of their mass function
with redshift. Although the satellites often form groups, they are loosely bound within them and do not interact with
each other. The infall of a large group could however explain the observed peculiar distribution of the Local Group
satellites, but only if it occurred recently. Mergers between prospective subhaloes are significant only during an early
stage of evolution, i.e. more than 7 Gyr ago, when they are still outside the host haloes. Such events could thus
contribute to the formation of more distant early type Milky Way companions. Once the subhaloes enter the host halo the
mergers become very rare.

\end{abstract}


\section{Introduction}

According to the favoured $\Lambda$CDM model galaxies are formed in a hierarchical way (White \& Rees 1978).
Small galaxies form first and then merge into larger structures. In this scenario the
surviving dwarf galaxies can be viewed as tracers of the early Universe. Understanding their properties and
evolution is crucial to understanding the $\Lambda$CDM Universe itself.

The Local Group (LG) is the closest and best studied object of the extragalactic scale (see van den Bergh 1999
for a review). Apart from the two large spirals, the Milky Way (MW) and Andromeda (M31), it is populated by
several tens of dwarf galaxies which by their morphological properties can be divided into irregulars,
ellipticals and spheroidals (Mateo 1998). They are known as the classical dwarfs. Recently we have
witnessed many new discoveries of ultra faint satellite galaxies in the halo of the MW (e.g. Sakamoto
\& Hasegawa 2006; Zucker et al. 2006; Belokurov et al. 2008). The dynamical properties of these objects are
still very poorly studied, but their alleged extremely high mass to light ratios (Simon \& Geha 2007) suggest
that they might be a rather different class of objects.

\begin{figure}
\begin{center}
\leavevmode
    \epsfxsize=7.2cm
    \epsfbox[0 0 490 460]{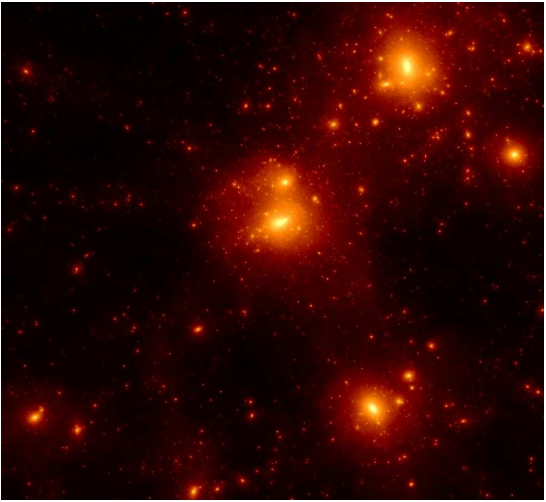}
\end{center}
    \caption{Overview of the simulated LG. M31 is located in the centre of the image, MW at the top right.
	The image size is 3 $h^{-1}$Mpc.}
\label{LG}
\end{figure}

One could expect that an early Universe progenitor of a dwarf galaxy should be similar for all types of the
present classical dwarfs (Kravtsov, Gnedin \& Klypin 2004) and would consist of a dark matter halo of mass
around $10^9$ M$_\odot$. This amount of mass would allow it to accrete gas even in the highly ionized
environment of the early Universe. The conservation of angular momentum would require that the gas and the
stellar component formed a disk rather than a spheroid. Such an object would then undergo evolution due to baryonic
processes such as cooling, star formation, supernova feedback, photoevaporation of the gas and heating by the cosmic
ultraviolet background. It remains to be seen if these or rather the environmental, purely gravitational processes
were the dominant factor in shaping dwarf galaxies. In this work we focus on the latter which can be modelled with
simulations following only the dark matter component.

The studies of stellar populations and kinematics of dwarf galaxies provide some general hints (see Tolstoy, Hill \&
Tosi 2009 for a recent review) concerning the possible environmental effects. The late type dwarfs (dwarf irregulars)
possess rich star formation histories, high amounts of hydrogen in stellar disks and significant angular momenta. They
are usually isolated objects (except for the Magellanic Clouds). These facts suggest that they evolved in isolation
only by the internal baryonic processes. On the other hand, the early type dwarfs usually lack gas and
do not show any signs of recent star formation. They also possess very low or no angular momentum. They typically are
close companions to the MW and M31, but even those isolated ones like Tucana or Cetus could have evolved in the
vicinity of the large spirals and been ejected recently to more distant orbits (Gill, Knebe \& Gibson 2005; Sales et
al. 2007a; Ludlow et al. 2009). They have high mass-to-light ratios from about ten to a few hundred solar units. They
could have evolved from a disk like those present in dwarf irregulars by strong gravitational interactions. These
interactions could take the form of mergers or tidal forces.

The latter idea has been developed into the so-called tidal stirring scenario (Mayer et al. 2001). Using
N-body simulations it has been demonstrated that indeed a transformation from a disk to the dwarf spheroidal
is possible by pure tidal interactions with a host galaxy. Mayer et al. (2001) have shown that low surface
brightness disks would produce dwarf spheroidal galaxies we observe today, while high surface brightness disks
would lead to the formation of dwarf ellipticals. Depending on the adopted star formation history this
scenario is able to reproduce dwarf spheroidals with moderate mass-to-light ratios (Klimentowski et al.
2007, 2009) as well as strongly dark matter dominated ones if the gas is expelled early on (Mayer et al.
2007).

Mergers and interactions between dwarfs could be another channel for the formation of early type dwarfs
since they could lead to a very strong evolution  (e.g. Knebe et al. 2006; Angulo et al. 2009), as in the case of larger
galaxies (e.g. Springel, Di Matteo \& Hernquist 2005). It has also recently been proposed that dwarf galaxies might be
accreted on to their hosts in groups which may explain e.g. the particular distribution of dwarfs around the MW and M31
(Libeskind et al. 2005; Metz, Kroupa \& Jerjen 2007, 2009a; D'Onghia \& Lake 2008; Li \& Helmi 2008; Metz et al.
2009b). In this work we discuss this second scenario of the formation of early type dwarf galaxies using a cosmological
$N$-body simulation of a LG. Our purpose is to study mergers and interactions of haloes which end up as subhaloes. We
also consider the infall of these objects together as a group of small haloes.

The paper is organized as follows. Section 2 contains the description of the simulation used in this analysis and the
halo finding algorithm which provides the basis for this study. In section 3 we characterize the main properties of the
subhalo population of the two largest haloes; we discuss their mass functions, their survival times and the
evolution of their masses. In section 4 we study the behaviour of subhaloes in groups; we provide the
statistics of groups around the two largest haloes, describe the mass functions of the largest groups and
follow their history. The effect of mergers and interactions between subhaloes is discussed in section 5. The discussion
follows in section 6.

\begin{figure*}
    \leavevmode
    \epsfxsize=7cm
    \epsfbox[-20 0 350 380]{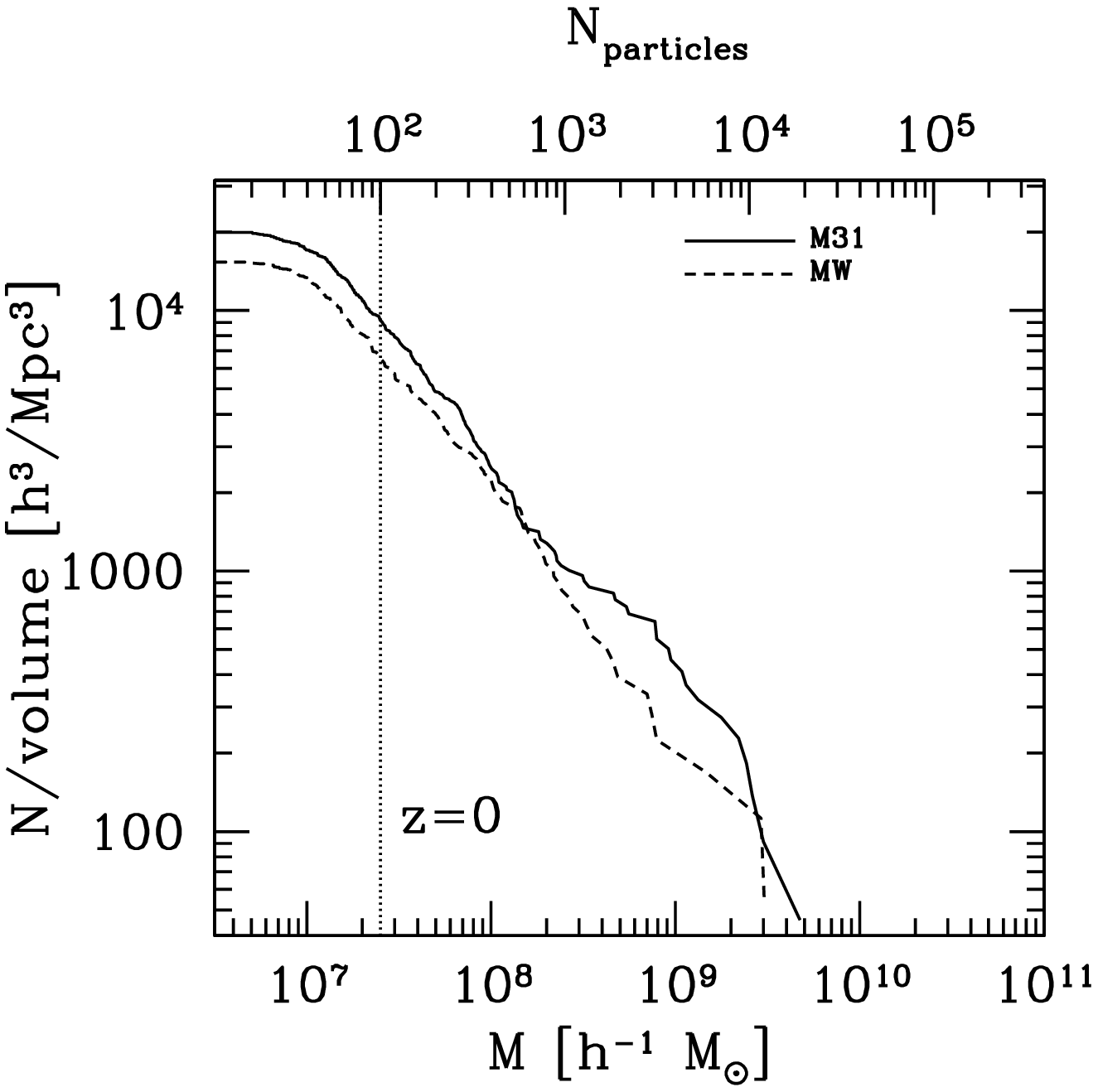}
    \epsfxsize=7cm
    \epsfbox[-20 0 350 380]{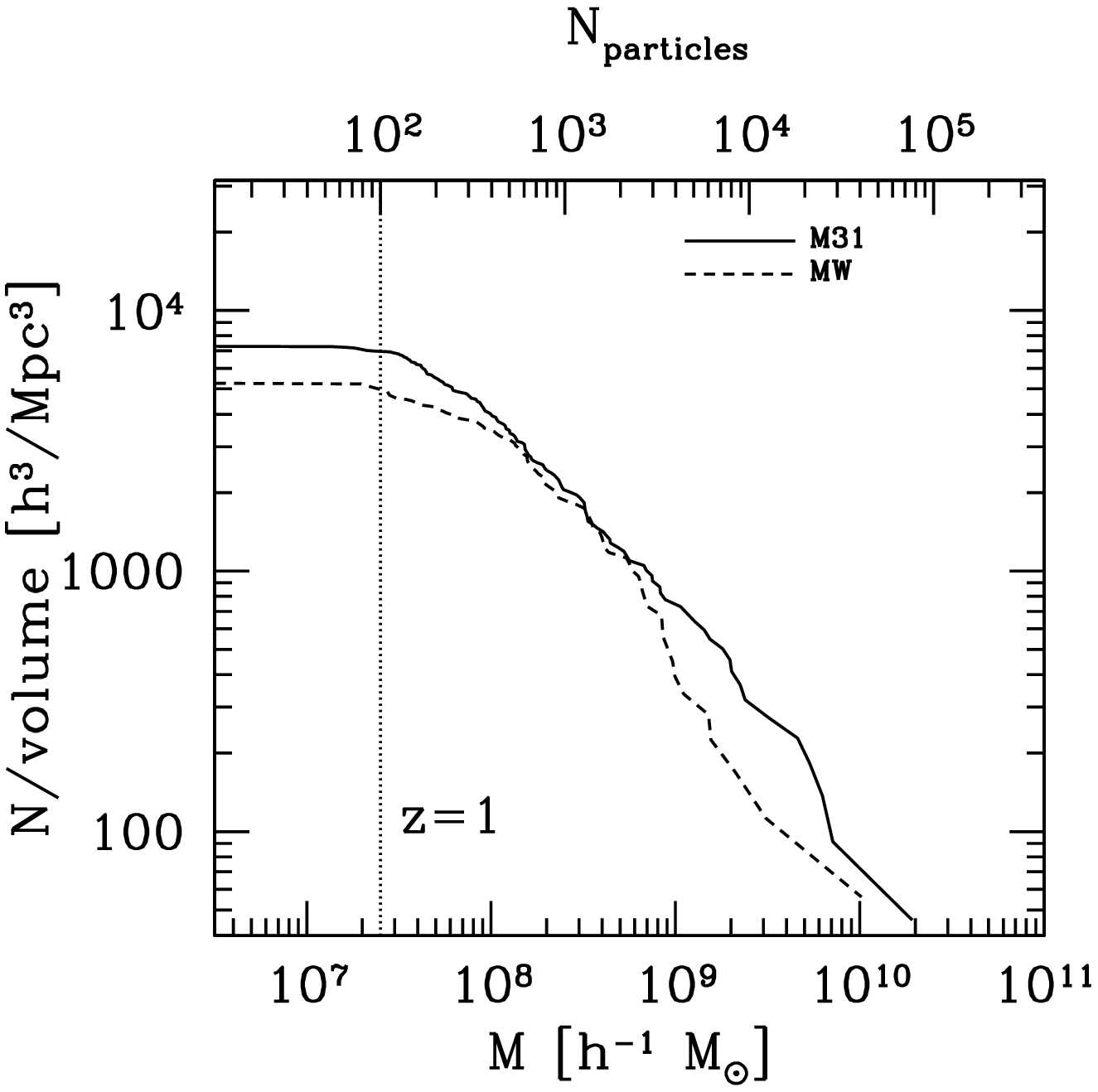}
    \caption{The cumulative mass function of subhaloes found within one virial radius of two
	most massive haloes normalized by the volume of that region. The left panel shows $N(>M)$ at redshift $z=0$.
	The right panel presents the mass function of the same haloes traced back to redshift $z=1$.
	The upper axis indicates the number of particles in a halo. Vertical dotted lines indicate our completeness
	limit of 100 particles.}
\label{mf}
\end{figure*}

\begin{table}
\caption{The properties of the LG: distance and
relative velocity between the two main members, the velocity dispersion
around the Hubble flow, the overdensity within a sphere of 7 Mpc centred
on the LG, distance from the LG to the Virgo cluster and the virgocentric flow.
The second column gives the values from the simulation. For comparison, the third column provides also the
observed values. The references for the observed values in the last column are: 1 - Karachentsev et al. (2004); 2 - van
der Marel \& Guhathakurta (2008); 3 - Tikhonov \& Klypin (2009); 4 - Hudson (1993); 5 - Fouqu\'e et al. (2001); 6 -
Tonry et al. (2000).}
\centering
\begin{tabular}{cccc}
\hline
Property & Simulated & Observed  & Ref.\\
\hline
$d$ (M31--MW) 				  & 0.91 $h^{-1}$ Mpc   & 0.77 Mpc         & 1 \\
$V_{\rm rel}$ (M31--MW) 		  & -193 km s$^{-1}$    & -130 km s$^{-1}$ & 2 \\
$\sigma_{\rm H} (r < 7$ Mpc)    	  & 100 km s$^{-1}$     & 90 km s$^{-1}$   & 3 \\
$\delta\rho/\overline{\rho}\ (r < 7$ Mpc) & 0.82                & 0.8              & 4 \\
$d$ (LG--Virgo)  			  & 10.9 $h^{-1}$Mpc    & 18.0 Mpc         & 5 \\
Virgocentric flow 			  & -255 km s$^{-1}$    & -409 km s$^{-1}$ & 6 \\
\hline
\end{tabular}
\label{LGprop}
\end{table}

\begin{table*}
\caption{The properties of the five most massive haloes identified in
the simulation. The columns list the assigned
name of a halo, its virial mass, the virial radius, the number of
particles inside the virial radius and the spin parameter (according to
Peebles definition).}
\centering
\begin{tabular}{cccrcc}
\hline
Name & $M_{\rm vir} [10^{11}h^{-1}$M$_\odot]$ & $r_{\rm vir}[h^{-1}$kpc]
& $N\ \ \ \ $ & $\lambda$ \\
\hline
Andromeda (M31) & $5.69$ & $173.5$ & $2239732$ & 0.0686  \\
Milky Way (MW)  & $4.62$ & $162.0$ & $1821419$ & 0.0607   \\
3rd halo        & $2.66$ & $134.7$ & $1045831$ & 0.0722  \\
4th halo        & $2.24$ & $127.2$ & $883066$  & 0.0219  \\
5th halo        & $2.01$ & $122.7$ & $738875$  & 0.0420   \\
\hline
\end{tabular}
\label{five}
\end{table*}

\section{The simulation}

We analyze a constrained dark matter simulation of the LG.  This
simulation is part of the \clues project\footnote{{\tt
http://clues-project.org}}, a collaboration whose main goal is to
produce realistic cosmological simulations of the Local Universe by
imposing observational constrains on the mass and velocity fields of
the initial random Gaussian fluctuation realizations. For this
simulation we have used a box of $64 \hMpc$ size assuming a spatially
flat cosmological model with WMAP3 parameters (Spergel et al. 2007):
$\Omega_{\rm m} = 0.24$, $\Omega_{\rm b} = 0.042$, $\Omega_{\Lambda} =
0.76$, the Hubble constant $h=0.73$, the normalization $\sigma_8 =
0.75$ and the slope $n=0.95$ of the power spectrum.

First, a
constrained density field on a grid of $256^3$ mesh points was
obtained applying the Hoffman \& Ribak (1991) algorithm
for generating constrained realizations of Gaussian random fields. As observational constraints
we have used the radial velocities of galaxies drawn from the MARK III catalogue
(Willick et al. 1997), Surface Brightness Fluctuation survey (Tonry et al. 2001) and the local volume
galaxy catalog (Karachentsev et al. 2004) as well as the positions of
nearby X-ray selected clusters of galaxies (Reiprich \& B{\"o}hringer
2002). The algorithm has been described in detail in Zaroubi, Hoffman
\& Dekel (1999), Kravtsov, Klypin \& Hoffman (2002) and Klypin et
al. (2003). With this algorithm to calculate the initial conditions
the resulting simulation contains the main features which characterize
the Local Universe. In the large simulation box the Virgo as well
as the Coma cluster and the Great Attractor are approximately at the right
positions whereas the small scale structure is essentially random. A smaller
box like the one we discuss here contains an object which can be
identified as the Virgo cluster.

Within our simulation box we represent the linear power spectrum at
redshift $z=100$ by $N_{\rm max}=4096^3$ particles of mass $m_{\rm DM}
= 2.5 \times 10^{5} h^{-1} \Msun $. We then Fourier transform the
constrained density field and substitute the overlapping Fourier modes
in our otherwise random realization. At first we degrade the mass
resolution to $256^3$ particles and identify the position of the
simulated LG at $z=0$. To this end we start at the position of the
simulated Virgo cluster and search for a Local Group like object at
the right position. After identifying such an object we find all
the particles within a sphere of radius $2 \hMpc$ centred on the
simulated object and determine the Lagrangian coordinates of these
particles in the initial conditions.

In a next step we resimulate the evolution of the LG using the full
resolution ($4096^3$) within this sphere of radius $2 \hMpc$.  Here we
follow the algorithm described in Klypin et al. (2001) and degrade the
mass and force resolution in those areas that are far away from the
Lagrangian region from which the LG forms.  To this end we put
concentric regions around the high-resolution area, each of them
populated with particles 8 times more massive. In the end we have
five different mass refinements ranging from $4096^3$ at the
high resolution area to $256^3$ in the outer parts of the simulation
box. Thus we simulate with very high resolution the evolution of the
LG in the right environment. The evolution of the same region has been
also simulated including gas dynamics and star formation
(Libeskind et al. 2009).

The simulation has been performed using the TreePM parallel $N$-body
code GADGET2 (Springel 2005). For the high mass resolution particles
we used a fixed comoving Plummer equivalent softening of 500 $h^{-1}$
pc at early redshift and changed to 100 $h^{-1}$ pc physical since
$z=4$. For the rest of the mass refinement levels, we increase the
Plummer softening by a factor of approximately two at every level. To
follow the evolution of the LG we have stored in total 134 outputs
equally spaced in time, which translates into a time difference of 0.1
Gyr between consecutive snapshots. The overview of the simulated LG in
the final output is shown in Fig.~\ref{LG} and its properties are
listed in Table~\ref{LGprop}. The properties of its five most massive haloes
are given in Table~\ref{five}.

In order to identify haloes and subhaloes in our simulation we have
run the MPI+OpenMP hybrid halo finder \texttt{AHF} (\texttt{AMIGA}
halo finder, to be downloaded freely from
\texttt{http://www.popia.ft.uam.es/AMIGA}). \texttt{AHF} is an
improved version of the \texttt{MHF} halo finder (Gill, Knebe \&
Gibson 2004a), which locates local overdensities in an adaptively
smoothed density field as prospective halo centres.  The local
potential minima are computed for each of these density peaks and the
gravitationally bound particles are determined. We stress that our
halo finding algorithm automatically identifies haloes, subhaloes,
subsubhaloes, etc. For more details on the mode of operation and the
actual functionality we refer the reader to Knollmann \& Knebe (2009)
where the  \texttt{AHF} halo finder is described in detail.

For each halo, we compute the virial radius $r_{\rm vir}$ at which the
density $M(<r)/(4\pi r^3/3)$ drops below $\Delta_{\rm vir}\rho_{\rm
b}$ where $\rho_{\rm b}$ is the cosmological background density.  The
threshold $\Delta_{\rm vir}$ is computed using the spherical top-hat
collapse model and is a function of both cosmological model and time
(e.g. {\L}okas \& Hoffman 2001). For the cosmology we are using,
$\Delta_{\rm vir}=355$ at $z=0$. Subhaloes are defined as haloes which
lie within the virial region of a more massive halo, the so-called
host halo.  As subhaloes are embedded within the density of their
respective host halo their very own density profile usually shows a
characteristic upturn at a radius $r_{\rm t} \lesssim r_{\rm vir}$,
where $r_{\rm vir}$ would be their actual (virial) radius if they were
found in isolation.\footnote{Note that the actual density
profile of subhaloes after the removal of the host's background drops
faster than for isolated haloes (e.g. Kazantzidis et al. 2004a); only
when measured within the background still present we will find the
characteristic upturn used here to define the truncation radius
$r_{\rm t}$.} We use this `truncation radius' $r_{\rm t}$ as the outer
edge of the subhalo and hence (sub)halo properties (i.e. mass, density
profile, velocity dispersion, rotation curve) are calculated using the
gravitationally bound particles inside either the virial radius
$r_{\rm vir}$ for a host halo or the truncation radius $r_{\rm t}$
for a subhalo.

Once the halo finding is completed all haloes are traced back in
time. For this purpose the halo must be linked to its progenitor in
the previous simulation output. We do that using the following
prescription. A halo progenitor is identified in the previous snapshot
by maximizing the ratio $C_i^2/(N_i N_j)$, where $C_i$ is the number
of common particles shared between the $i$-th halo of the current
snapshot and the $j$-th halo of the previous snapshot while $N_i$ and
$N_j$ are the total numbers of particles in these haloes.

This simple formula works surprisingly well in identifying the correct
halo in the past. Unfortunately, sometimes the correct halo is missing
in the previous snapshot because it was not found by the halo
finder. This happens, for example, when a smaller halo passes close to
the centre of a larger one and due to its low density contrast the
halo finder may not be able to identify it. To take care of this
problem (wrong identification would lead to something which looks like
a halo splitting in two) we also check that the mass of the progenitor
is close to the mass of the descendant: if the mass ratio is smaller
than $0.8$ we look for the correct progenitor halo two (instead
of one) snapshots earlier. This parameter was tuned to the actual properties of the halo
finder. This procedure is applied
recursively until we find a credible progenitor in one of the higher
redshift snapshots. In practice, however, we never have to skip more
than one or two snapshots.

\begin{figure}
    \leavevmode
    \epsfxsize=7.2cm
    \epsfbox[0 10 340 370]{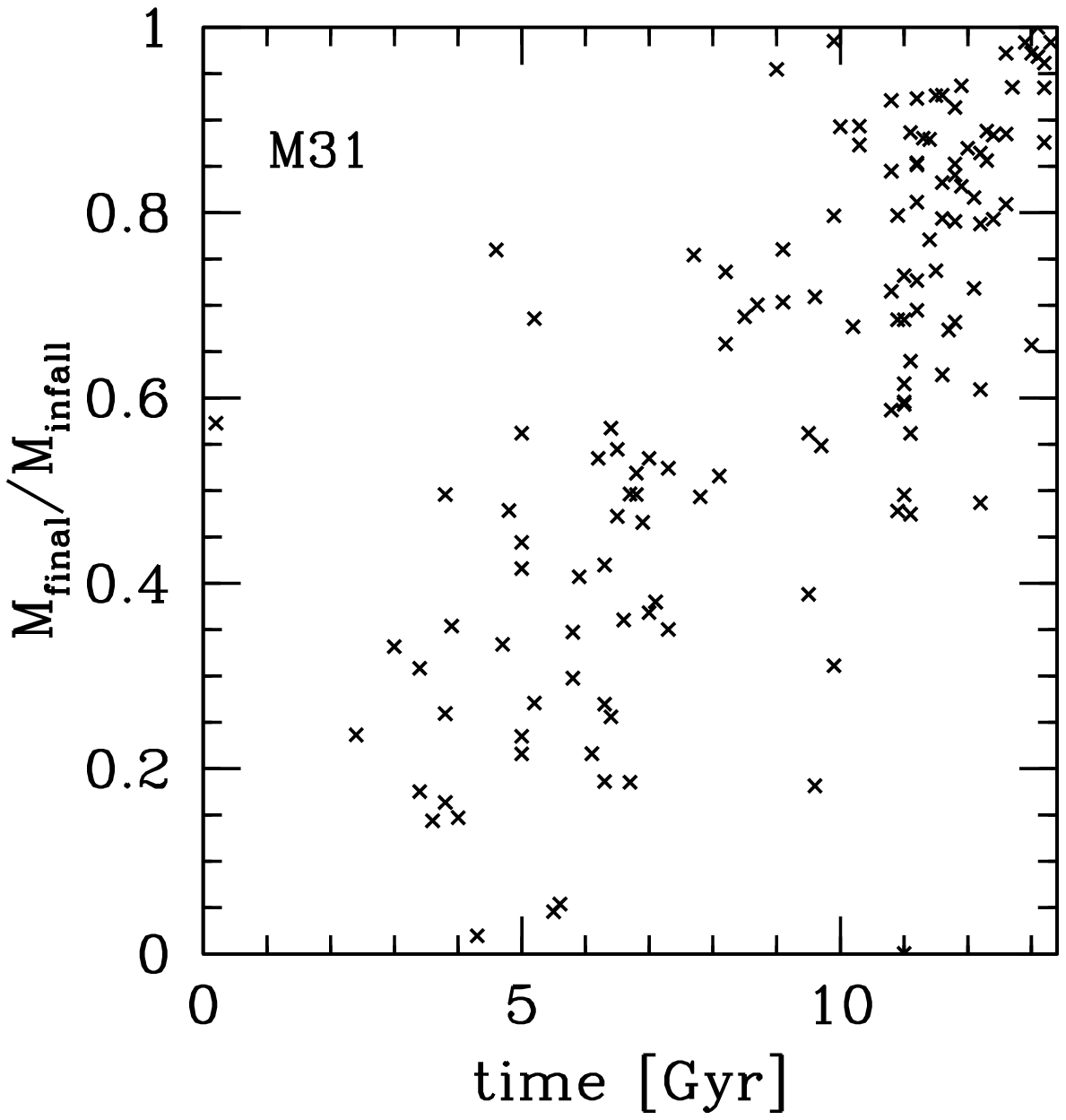}
    \epsfxsize=7.2cm
    \epsfbox[0 10 340 370]{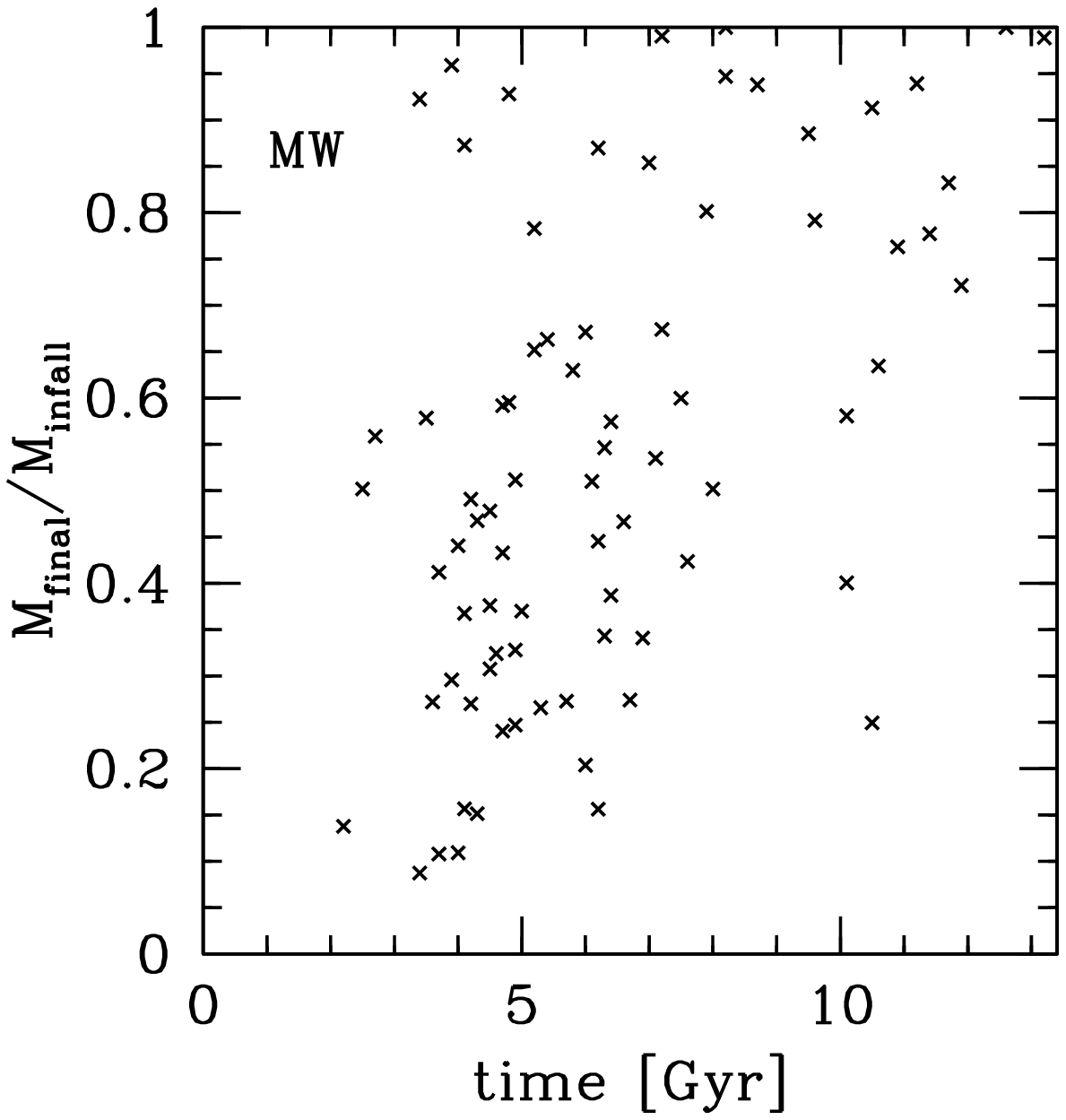}
\caption{The ratio between the final mass of a subhalo at the end of the simulation and its mass at the
infall time, as a function of infall time for M31 (upper panel) and MW (lower panel).}
\label{mchange}
\end{figure}

\begin{figure*}
    \leavevmode
    \epsfxsize=15cm
    \epsfbox{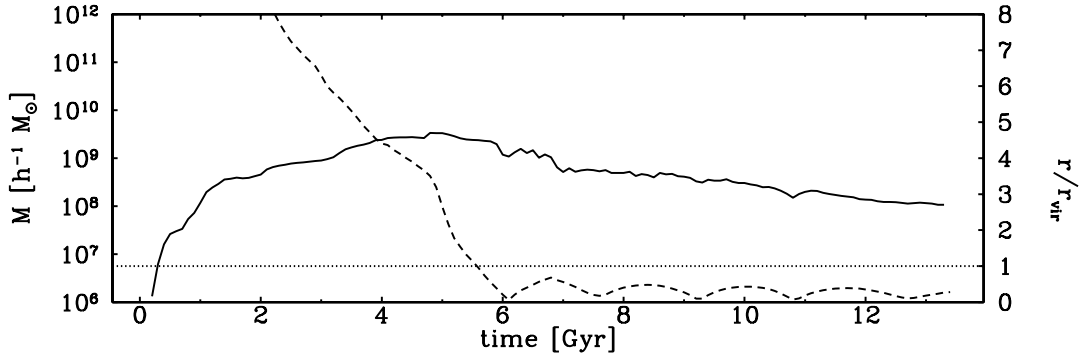}
\caption{Example of the evolution of a satellite subhalo. The solid line shows the mass of the subhalo and
the dashed one its distance from the host galaxy divided by its evolving virial radius, as a function of time. The dotted
line shows the distance of one virial radius from the host, which we adopt as a measure of the orbit entry point for the
subhalo. During the first stage of the evolution the subhalo gains mass. After entering the orbit around its host it starts losing
mass due to tidal stripping.}
\label{evolution}
\end{figure*}

\section{Properties of the satellites}

The first two rows of Table~\ref{five} list the properties of the two most
massive haloes identified in the last simulation output corresponding to redshift $z=0$. We will call the largest one
Andromeda (M31) and the second largest the Milky Way (MW). In this section we study in detail the properties of the
satellites of these largest haloes. The left panel of Fig.~\ref{mf} shows the
mass function $N(>M)$ of all subhaloes belonging to the two most massive haloes at redshift $z=0$. The subhaloes were
identified within one virial radius of their host. The relation was scaled by the volume of each region in which subhaloes
were selected. In this case the mass functions are very similar for both haloes.
Note that the shapes of the mass functions flatten towards smaller masses
signifying loss of completeness. This happens at masses corresponding to about 100 bound
particles or $2.5 \times 10^7$ $h^{-1}$ M$_\odot$. Unless otherwise stated, from now on we will consider only haloes
with masses above this value.

The right panel of Fig.~\ref{mf} shows the cumulative mass function of all subhaloes found at redshift $z=0$ traced back
to $z=1$. Comparing the panels we see that with decreasing redshift the number
density of small mass haloes increases and the whole slope of the relation steepens. This suggests that
subhaloes which were originally more massive lose mass during their evolution. This process is due to the
tidal interactions of subhaloes with their host halo and is common for satellites of all large galaxies (cf. Mayer
et al. 2001; Gill et al. 2004b; Giocoli, Tormen \& van den Bosch 2008).

This phenomenon is further illustrated in Fig.~\ref{mchange} which shows the ratio between the final mass of a
surviving satellite at the end of the simulation and its mass at the infall or orbit entry time as a function of
infall time, i.e. the time it crosses the virial radius of the host for the first time. The tendency for mass loss is
clearly seen as the ratios are typically much lower than one especially for those haloes which became satellites early.
Figure~\ref{evolution} shows this process in detail for a single example subhalo.
During its evolution it first gains mass as the hierarchical formation
scenario predicts, but then reaches a point at which it starts losing mass. As expected, the moment when the
mass trend reverses is close to the time at which the subhalo becomes a satellite of a larger halo, which
will be its host for the rest of the simulation. From that moment on the satellite is being stripped by tidal forces
of the host. The mass loss obviously depends on the infall time but also on other parameters like the orbit.
The presented example is a rather extreme case. According to Fig.~\ref{mchange} most haloes do not lose that much
mass. Also not all subhaloes follow this path of evolution. Many of them just fall into the host halo and merge with
it.

\begin{figure}
    \leavevmode
    \epsfxsize=7.2cm
    \epsfbox[0 0 340 370]{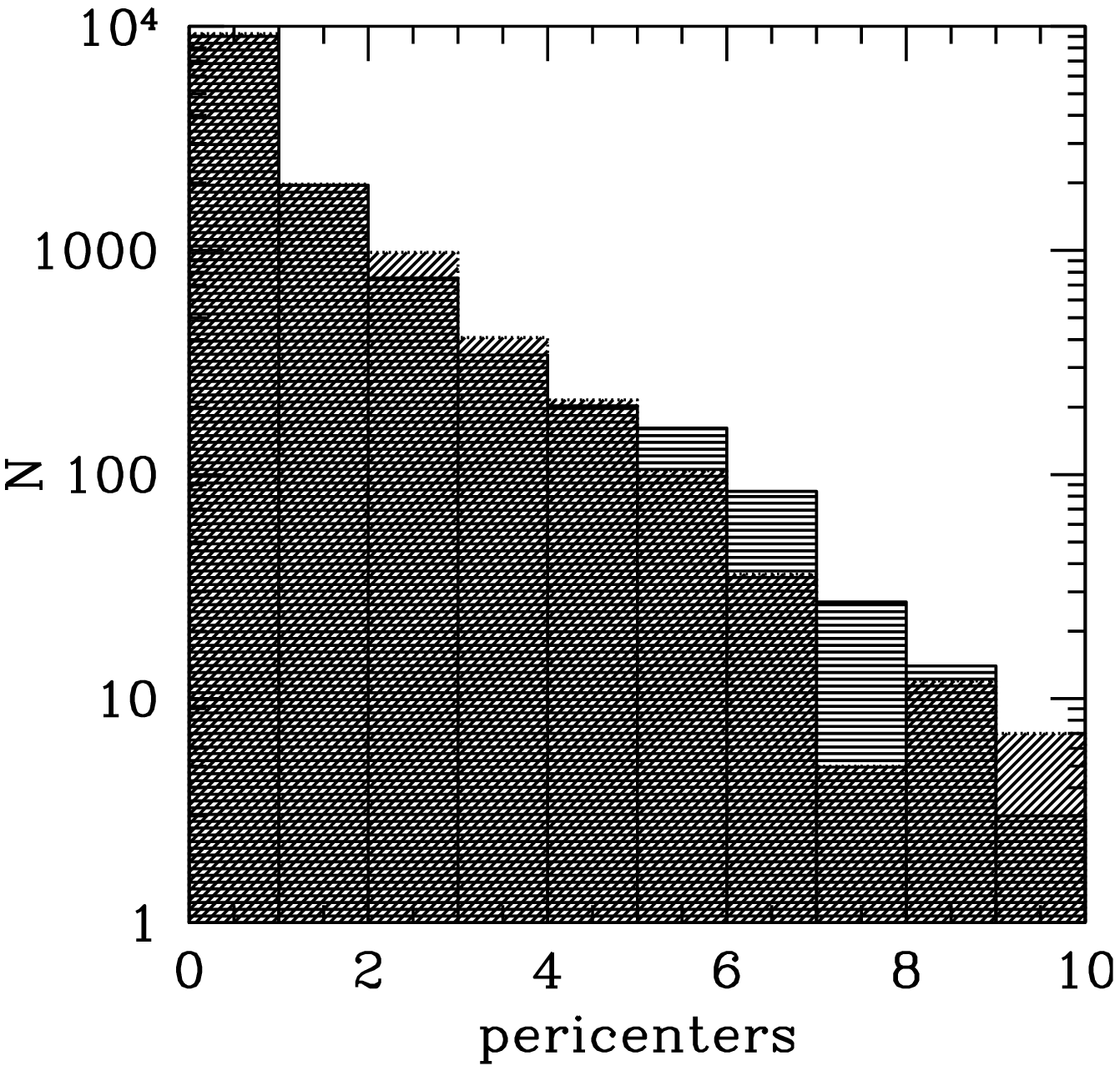}
    \epsfxsize=7.2cm
    \epsfbox[0 0 340 370]{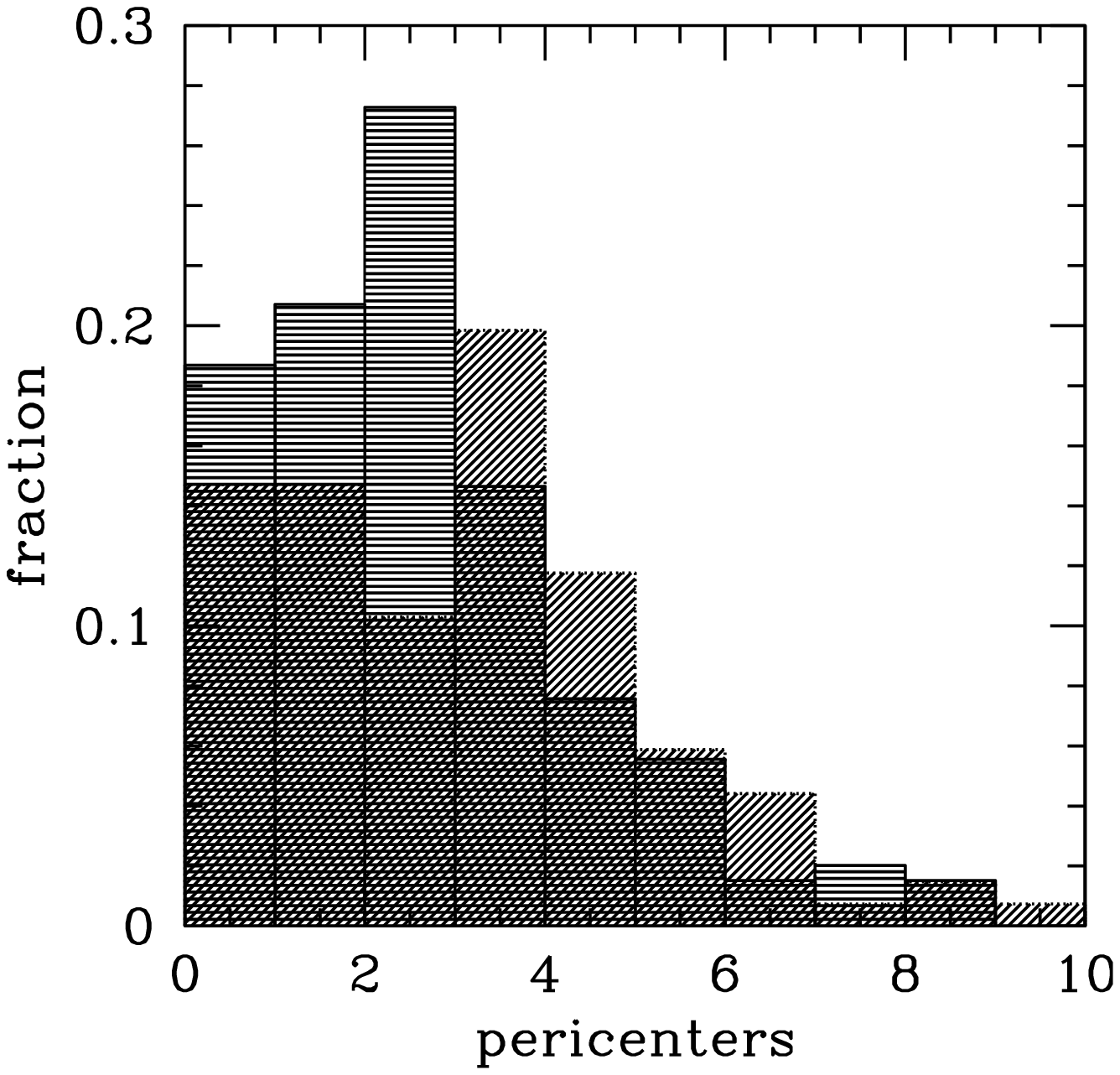}
\caption{The survival of subhaloes. The upper panel shows the numbers of pericentre passages survived by subhaloes
before their destruction
therefore it shows those which eventually {\em did not\/} survive until the present. Subhaloes with zero pericentre
passages were accreted directly on to the host halo and destroyed during their first passage. The lower panel
corresponds to those subhaloes which survived until the end of the simulation and shows the fraction of subhaloes that
survived a given number of pericentre passages to all surviving subhaloes separately in the M31 and in the
MW. Here subhaloes with zero pericentre passages have entered their host recently and have not yet reached
the pericentre of their orbit. In each panel the horizontal shading corresponds to the M31, the skewed one to
the MW.}
\label{pericentres}
\end{figure}
\begin{figure}
    \leavevmode
    \epsfxsize=7.2cm
    \epsfbox[0 0 340 370]{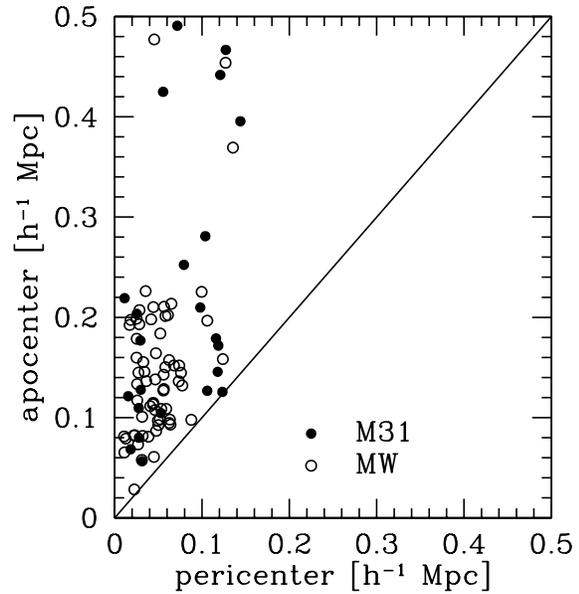}
\caption{The distribution of orbits for subhaloes that survived until the present, were found
inside the virial radius of the host and completed at least one orbit around it. The filled and open circles show the
apocentre versus pericentre distance respectively for M31 and MW subhaloes. The
line indicates the circular orbits.}
\label{lastorb}
\end{figure}

\begin{figure}
    \leavevmode
    \epsfxsize=7.2cm
    \epsfbox[0 0 370 370]{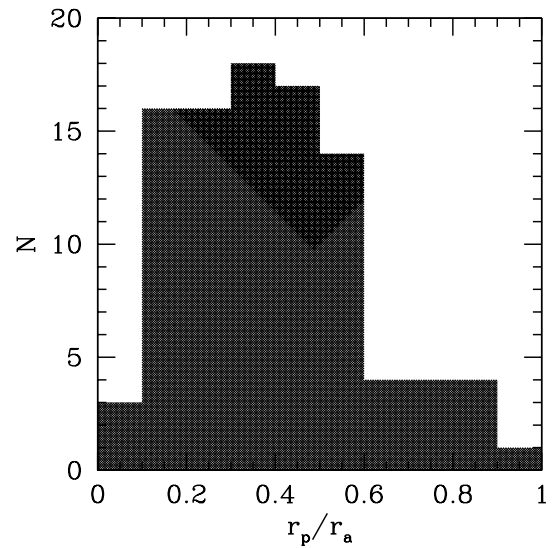}
\caption{The distribution of peri- to apocentre ratio for the same sample of M31 and MW subhaloes as shown in
Fig.~\ref{lastorb}.}
\label{histrat}
\end{figure}

Figure~\ref{pericentres} shows the survival time of a satellite in terms of the number of pericentre passages
it sustains. In the upper panel only those satellites were selected which {\em did not\/} survive until the end of the
simulation and were completely destroyed. Many subhaloes do not survive even a single pericentre passage, they fall
straight inside and merge with the host. This behaviour is in agreement with other $N$-body studies where it was found
that majority of stars in the present MW halo comes from the most massive subhaloes that were accreted in the past
(e.g. Bullock \& Johnston 2005; Sales et al. 2007b). There are however still more than a hundred of satellites per
halo, which were able to survive up to eight pericentre passages. Their orbital history is illustrated by the histogram
in the lower panel of Fig.~\ref{pericentres}. Presumably, these might be the dwarf spheroidal galaxies of the LG that
we presently see as, according to the tidal stirring scenario (Mayer et al. 2001; Klimentowski et al. 2009), a
significant number of pericentric passages accompanied by substantial mass loss are required to transform a disky
stellar component into a spheroid.

Actually, at the end of the simulation, M31 has 198 and the MW
136 surviving subhaloes inside the virial radius. It is interesting to look at the distribution of the shapes of the
orbits of these subhaloes. Figure~\ref{lastorb} plots the apocentre versus pericentre ($r_{\rm a}$ and
$r_{\rm p}$) for those subhaloes that survived until the end and completed at least one orbit around their host. Note
that M31 subhaloes are strongly underrepresented because many of them found inside the virial radius at the end
have not yet completed a single orbit (see the next section). Figure~\ref{histrat} presents the distribution of the
ratio $r_{\rm p}/r_{\rm a}$ of the same subhaloes. We can see that there is a strong preference for rather eccentric
orbits in agreement with the results found by Gill et al. (2004b) and Diemand, Kuhlen \& Madau (2007).

\begin{figure}
    \leavevmode
    \epsfxsize=7.2cm
    \epsfbox[-20 0 340 370]{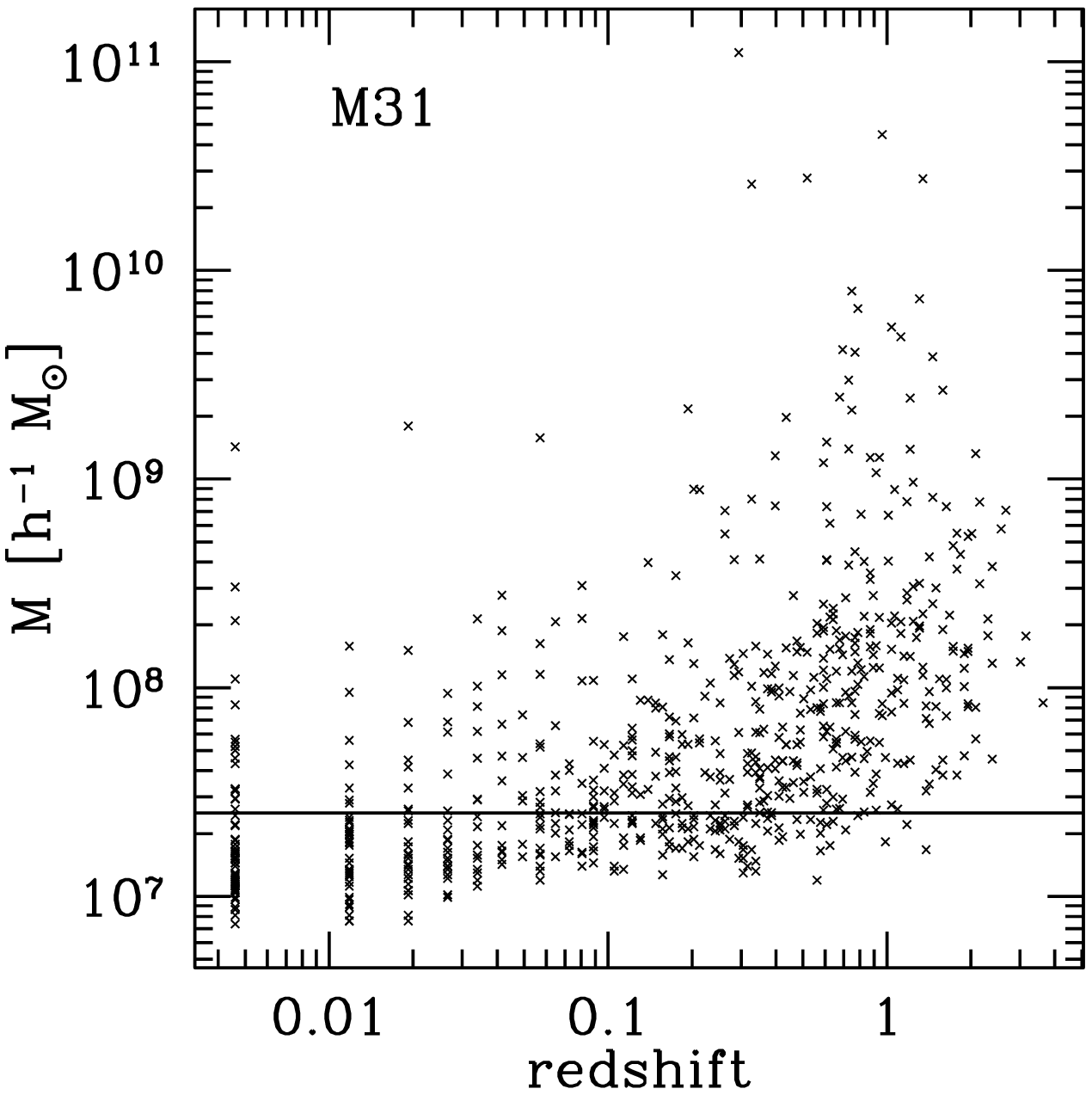}
    \epsfxsize=7.2cm
    \epsfbox[-20 0 340 370]{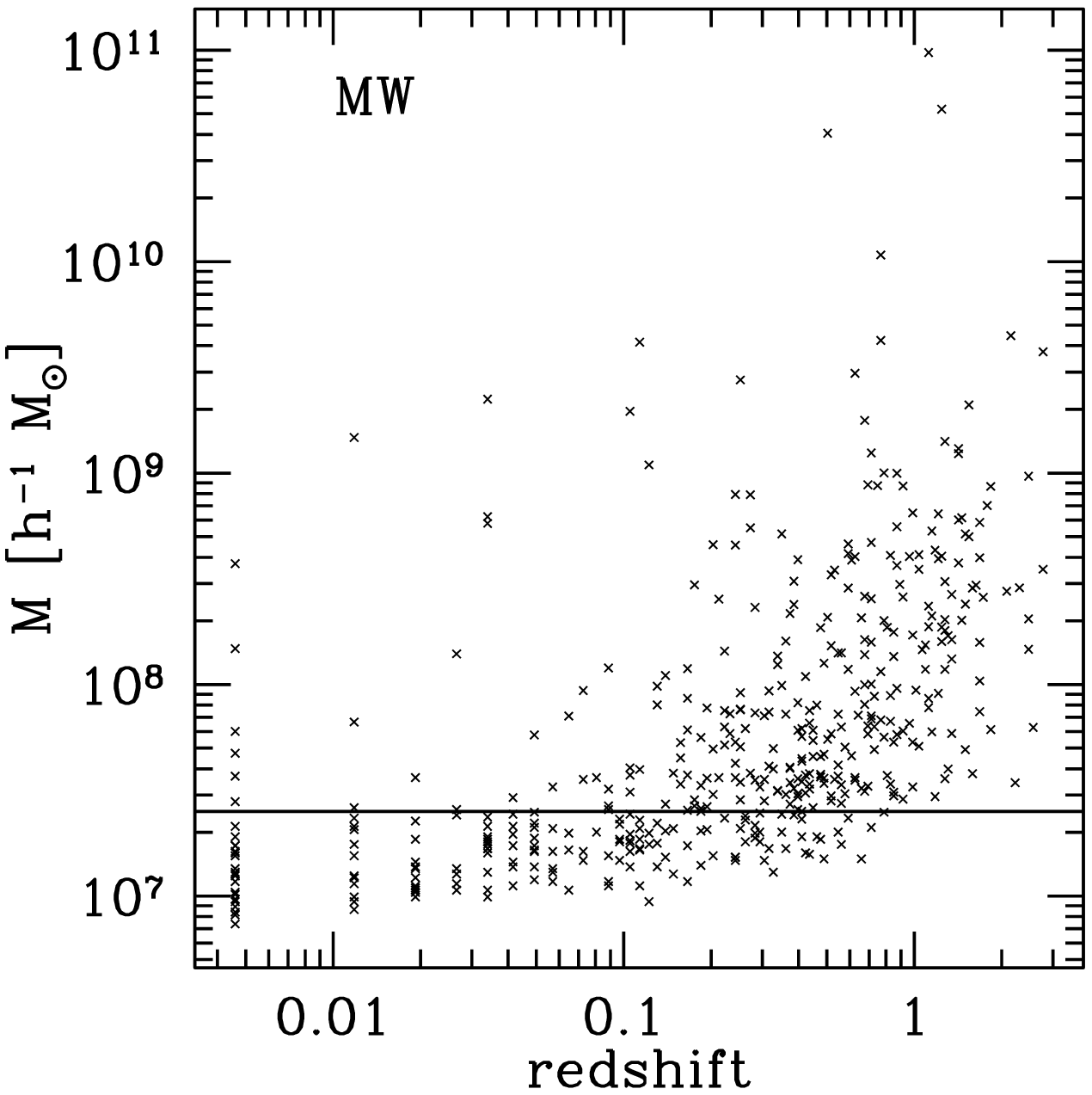}
\caption{The maximum mass of a subhalo as a function of redshift
at which this mass was reached. The mass corresponds to the moment when a subhalo enters its orbit
around the host halo and starts losing mass, as shown in Fig.~\ref{evolution}. The upper diagram
is for the subhaloes of M31 while the lower one for the subhaloes of the MW. Only subhaloes
which survived until the present time were plotted. The solid line in each panel indicates our completeness limit
of 100 particles, corresponding to $M=2.5 \times 10^7 h^{-1}$ M$_\odot$.}
\label{massrev}
\end{figure}

Figure~\ref{massrev} shows the mass of subhaloes at the moment of orbit entry as a function of redshift at
which the infall of the subhalo takes place. The Figure allows us to hypothesize on the possible evolution of the
subhaloes in different mass ranges. The majority of large mass subhaloes (with masses around $10^9$
M$_\odot$) enter their orbits at around $z=1$. These are most probably the progenitors of dwarf
spheroidal galaxies. As shown in Klimentowski et. al (2009) a halo of this mass possessing a stellar disk at
$z=1$ has enough time to form a dwarf spheroidal by the present by tidal stirring. The mass loss that
occurs in this process can be very significant as demonstrated by Fig.~\ref{mchange} (cf. Mayer et al. 2001; Hayashi et
al. 2003; Gill et al. 2004b; Kazantzidis, Moore \& Mayer 2004b; Kampakoglou \& Benson 2007; Klimentowski et al. 2007,
2009; Pe\~narrubia, Navarro \& McConnachie 2008; Giocoli et al. 2008) which at the end of the evolution leaves us with
masses of the order of few times $10^7 $M$_\odot$, as indeed measured for most dwarf spheroidals (e.g. {\L}okas et al.
2008; {\L}okas 2009).

The subhaloes that start orbiting their host later most probably correspond to dwarf irregular galaxies, as
they already have the correct masses around $10^8-10^9$ M$_\odot$ and not enough time to evolve anymore. If
they are also dwarf spheroidals today they would need to be dwarf spheroidals already at the moment of orbit
entry. A process which could lead to the formation of a dwarf spheroidal (or a dwarf elliptical) at large
distance from a host galaxy is a collision of its progenitor with another similar object.

Another class of satellites that can be distinguished in Fig.~\ref{massrev} are the galaxies which
enter their orbits early and possess low masses of the order of a few times $10^7$ M$_\odot$. They
would then evolve by tidal stirring, like normal dwarf spheroidals, leading to similar objects but of much
smaller masses. These could be the progenitors of the ultra faint dwarfs with expected present masses of
around $10^6$ M$_\odot$ and smaller. As demonstrated in Klimentowski et al. (2009) the tidal stripping
alone does not significantly change the mass-to-light ratios so if they indeed possess high $M/L$ values,
they must be due to the processes involving the evolution of the baryonic component, in particular the gas
dynamics. Indeed, their very low initial masses could suppress star formation in the early stages (e.g.
White \& Rees 1978; Kauffmann, White \& Guiderdoni 1993; Mayer et al. 2007). Then they would host very
few stars from the beginning which explains why they possess so few stars at the present time (Haiman,
Rees \& Loeb 1997).

Obviously, the mass resolution of the simulation described here is not sufficient to study the tidal
stripping of satellites in detail. Figures~\ref{mchange} and \ref{evolution} should therefore be
considered only as indicative that this process is indeed present. In order to make reliable predictions on
its effectiveness one needs to follow the evolution of dwarf galaxies with much higher resolution, which
usually means evolving a single dwarf in a fixed potential of the host, as done e.g. in Klimentowski et al.
(2007, 2009). Our purpose here is rather to investigate other possible scenarios for the formation of early
type dwarfs which involve neighbours. The following sections are therefore devoted to the evolution of
satellites in groups and the merging of satellites.

\section{Satellites in groups}

It has been recently proposed that the infall of subhaloes on to a large halo might occur in groups. The
claims have been based both on $N$-body simulations (e.g. Li \& Helmi 2008; Angulo et al. 2009) and
observational data (e.g. D'Onghia \& Lake 2008; Metz et al 2009a). The distribution of dwarf satellites of the MW also
suggests that they did not infall from completely random directions and form a disk-like structure (Metz et al. 2007,
2009a). If such groups of infalling haloes were bound the probabilities of their close
interactions and mergers before the infall would be much higher than for randomly distributed haloes. In this case the
haloes inside such groups could be the progenitors of at least some dwarf spheroidal or elliptical galaxies
forming them either by tidal interactions or collisions. In this section we study such a possibility in detail.

\begin{figure}
    \leavevmode
    \epsfxsize=7.2cm
    \epsfbox[0 0 340 370]{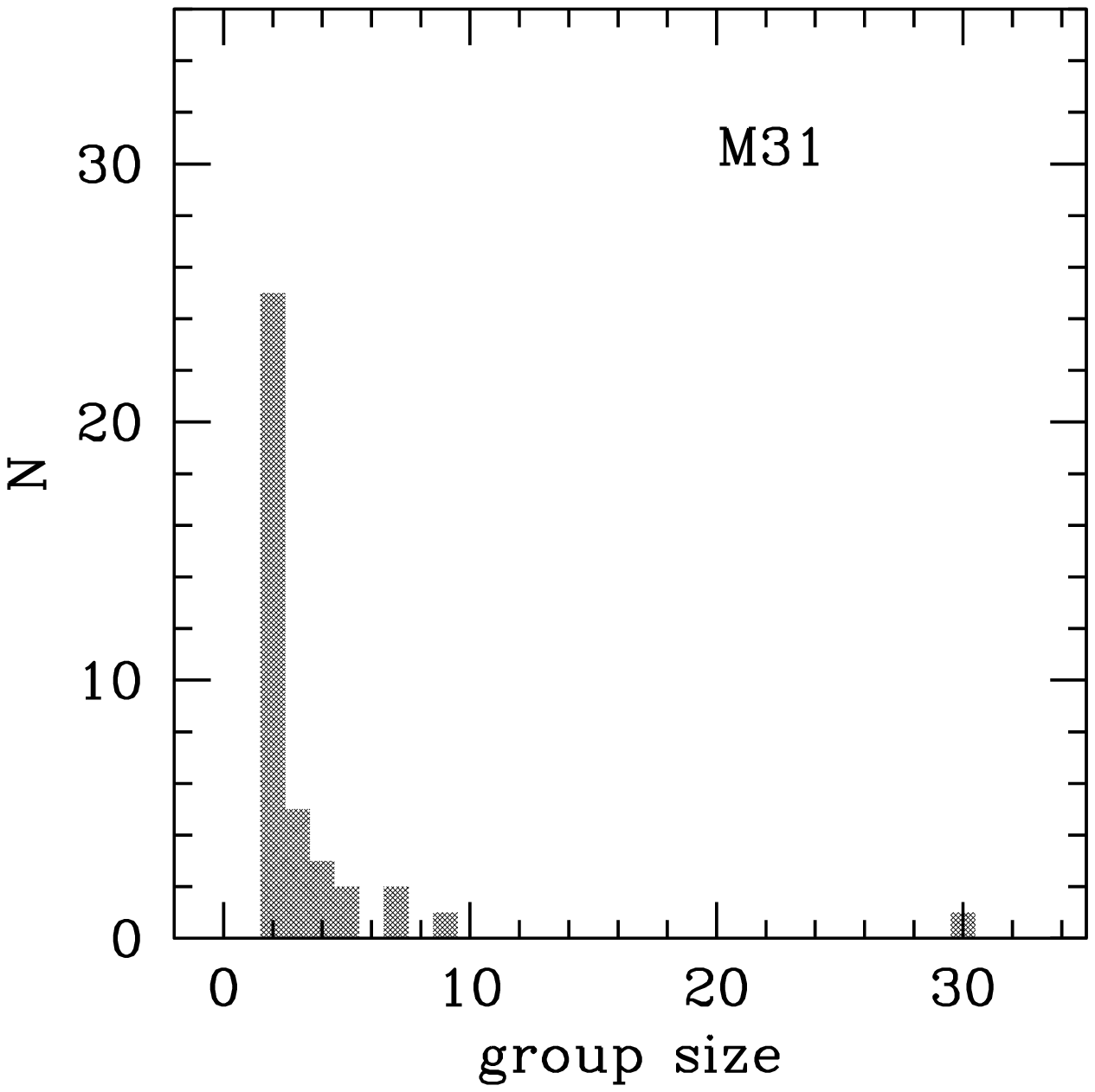}
    \leavevmode
    \epsfxsize=7.2cm
    \epsfbox[0 0 340 370]{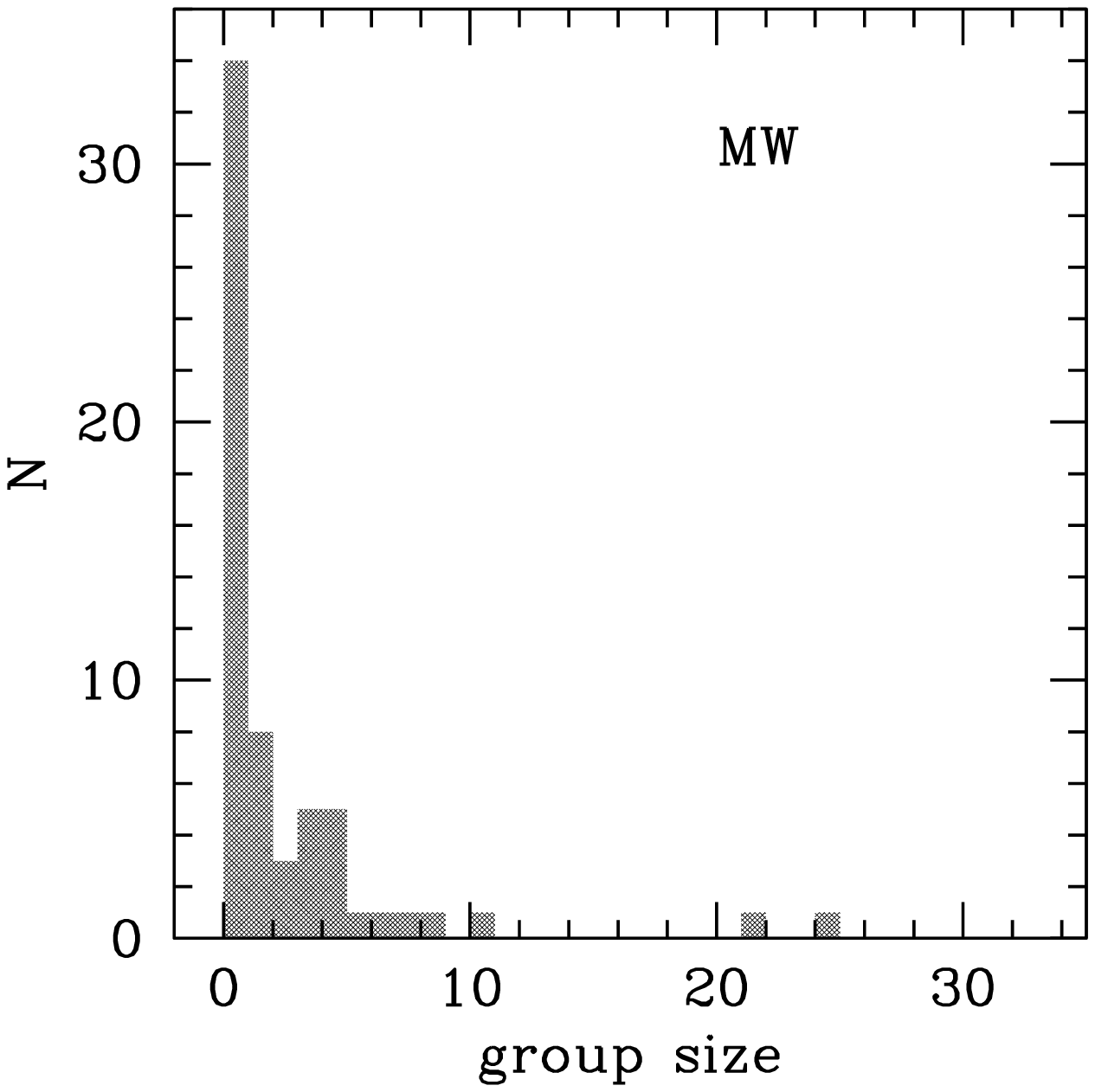}
\caption{The distribution of group size (measured as number of group members) in M31 (upper panel) and the MW
(lower panel). The groups were identified by applying the group finding algorithm with parameters: $l=100$ $h^{-1}$
kpc, $p=1.5$ Gyr, $t_1=7$ Gyr, $t_2=10.5$ Gyr for M31 and $l=100$ $h^{-1}$ kpc, $p=1.5$ Gyr, $t_1=0.2$ Gyr, $t_2=13.4$
Gyr for the MW. }
\label{histgroupsize}
\end{figure}

\begin{figure}
    \leavevmode
    \epsfxsize=7.2cm
    \epsfbox[0 0 340 370]{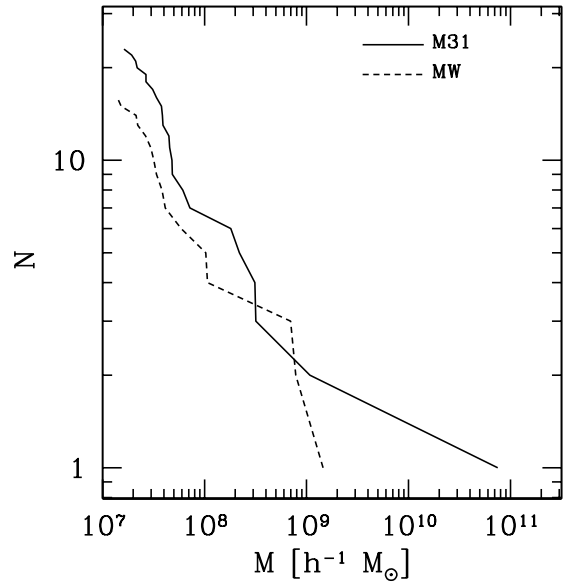}
\caption{The cumulative mass functions of the largest groups in M31 (LGA, solid line) and the MW (LGM,
dashed line).}
\label{mfgroups}
\end{figure}

\begin{figure}
\begin{center}
\leavevmode
    \epsfxsize=7.2cm
    \epsfbox[0 0 427 349]{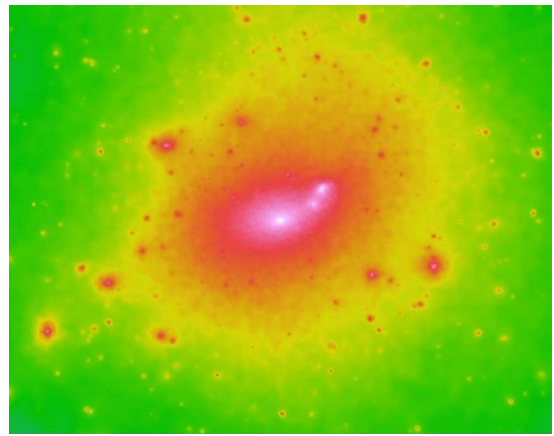}
\end{center}
    \caption{M31 together with its largest subhalo visible to the upper right of the main halo. The size of the image
is 200 $h^{-1}$kpc.}
\label{M31}
\end{figure}

\subsection{Group finding algorithm}

Our group finding algorithm is a generalized
version of the friends-of-friends (FOF) algorithm (Davis  et al. 1985). As in the classical FOF procedure,
we define a linking length. An additional parameter is the number of simulation outputs or a time period.
Two haloes are considered to form a group if they are closer to each other than the linking length $l$
for a given time period $p$. Each pair of haloes is checked. If one of the haloes which meet the criteria
is already a member of a group, then the other one is added. When both haloes are already members of
different groups, then the groups are merged. The algorithm reduces to the standard FOF when the period
is set to a single snapshot.

The algorithm is capable of finding many different kinds of groups. We have not used all its possibilities
though. One of the most important features of the algorithm is the fact that a group does not have to be
strictly defined in time. For example, two haloes could form a group at one time, while one of these haloes
could form a group with a third halo at a different time. Although we have two different groups of
two haloes they are linked together to form one group by a common halo. If we do not want such linking we
need to define a time period (with the starting moment $t_1$ and the ending moment $t_2$ where $t_2-t_1 \geq p$),
at which we want the algorithm to actually link the haloes. In practice, we use long time periods to
find the groups and study their history, and then reapply the algorithm with a short time period at the time
of infall to study the group behaviour in its host halo. The set of four parameters $l, p, t_1, t_2$
completely define the algorithm. Obviously, it can be run on different subsets of haloes. We have run the algorithm
for the subhaloes of the two most massive haloes, the M31 and the MW. A subhalo was defined as before as a halo which
in at least one simulation output was closer to the host than one virial radius.

\subsection{Results of the group finding for M31}

\begin{figure}
    \leavevmode
    \epsfxsize=6.5cm
    \epsfbox[0 -10 340 370]{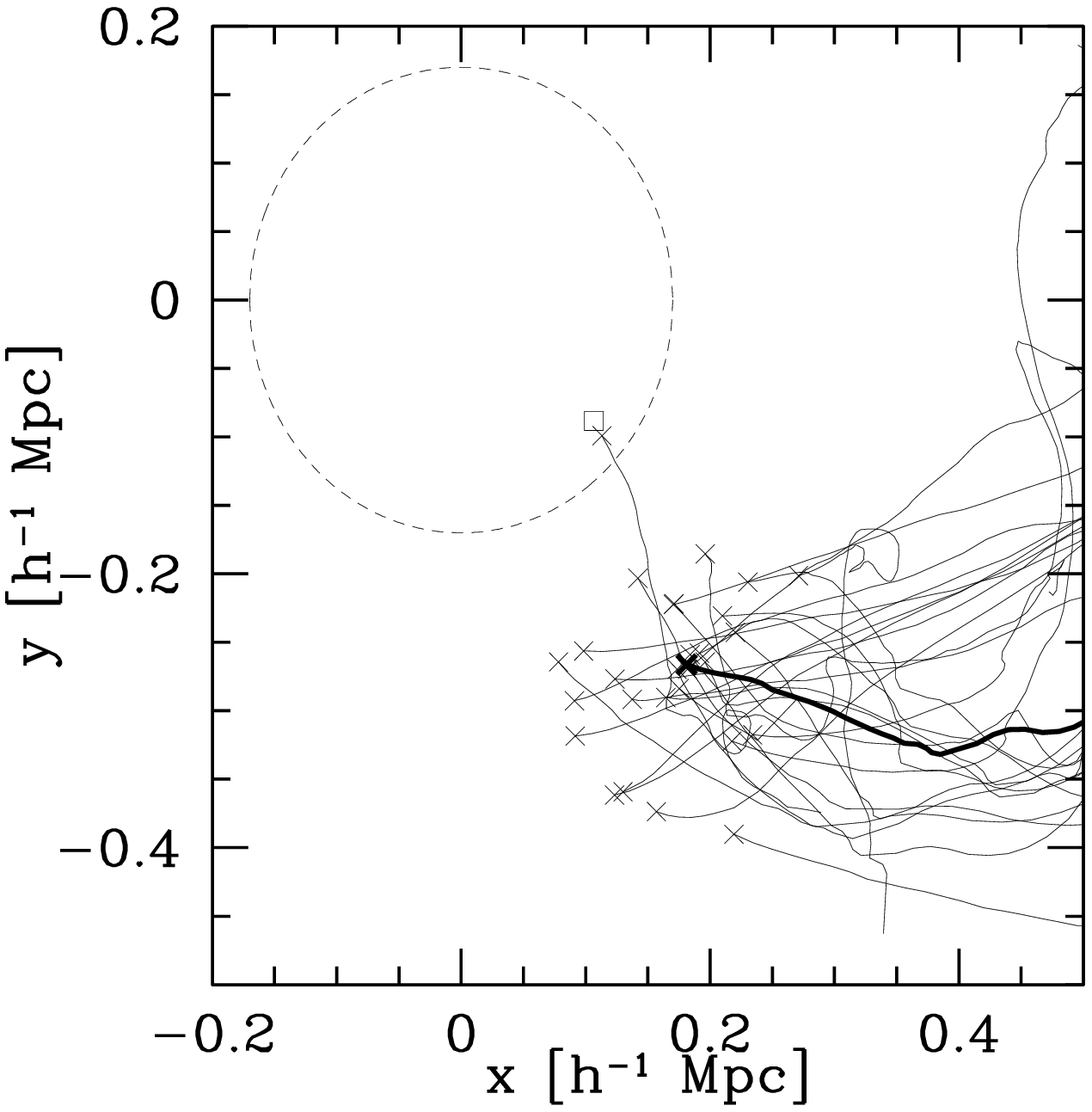}
    \epsfxsize=6.5cm
    \epsfbox[0 -10 340 370]{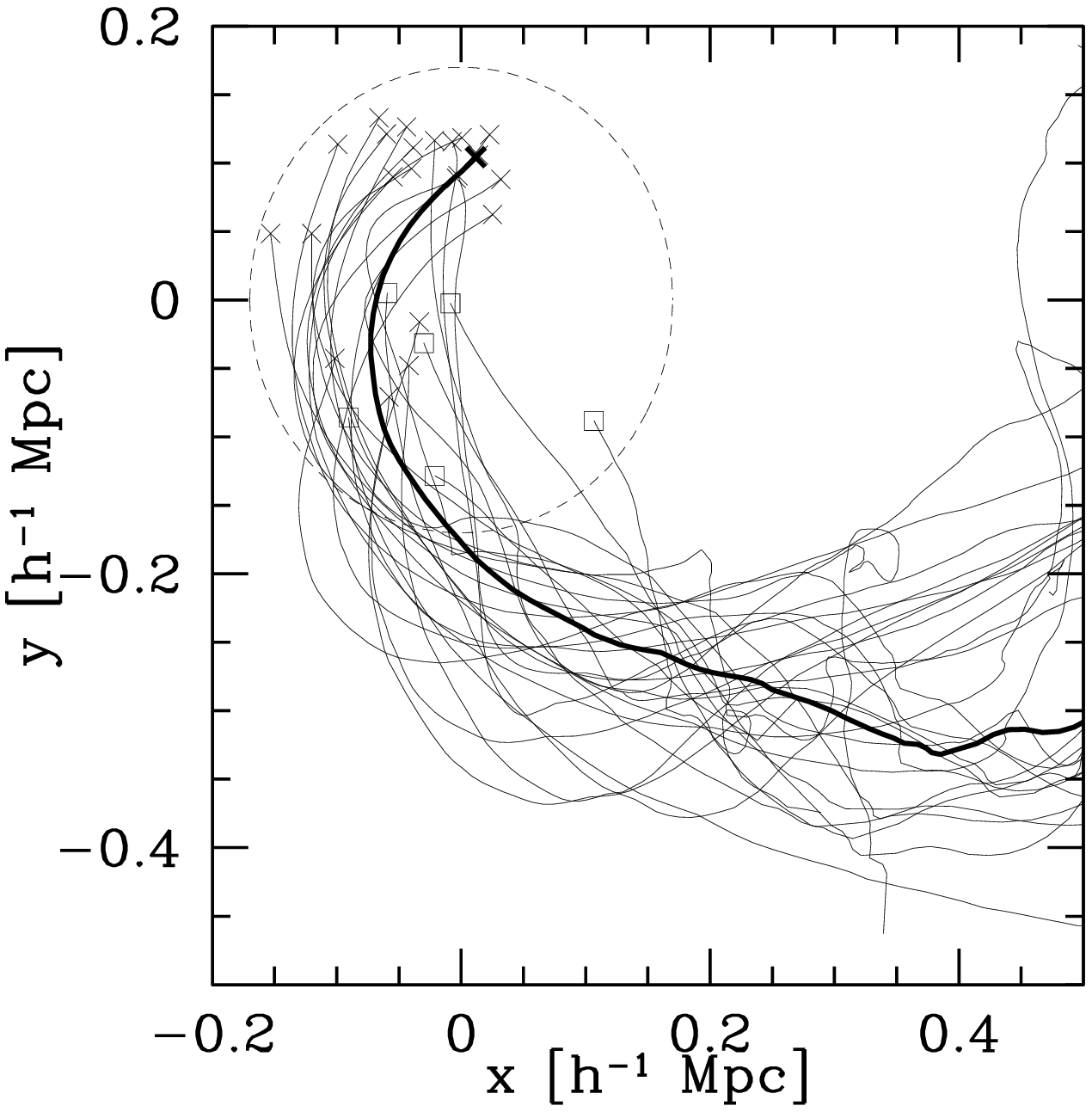}
\caption{Trajectories of the subhaloes in the Large Group of Andromeda (LGA) starting around 6 Gyr from the beginning
of the simulation. The upper panel shows the positions of
 subhaloes at 9 Gyr, the lower one at present (13.4 Gyr). In both panels the biggest
 halo is indicated by the thicker line. The
 dashed circle marks the present virial radius of M31. A square at the end of the line indicates that the halo was
destroyed before the end of the simulation, a cross shows the position of a halo which survived until the present.}
\label{lgapic}
\end{figure}

\begin{figure}
    \leavevmode
    \epsfxsize=7.2cm
    \epsfbox[-20 0 340 370]{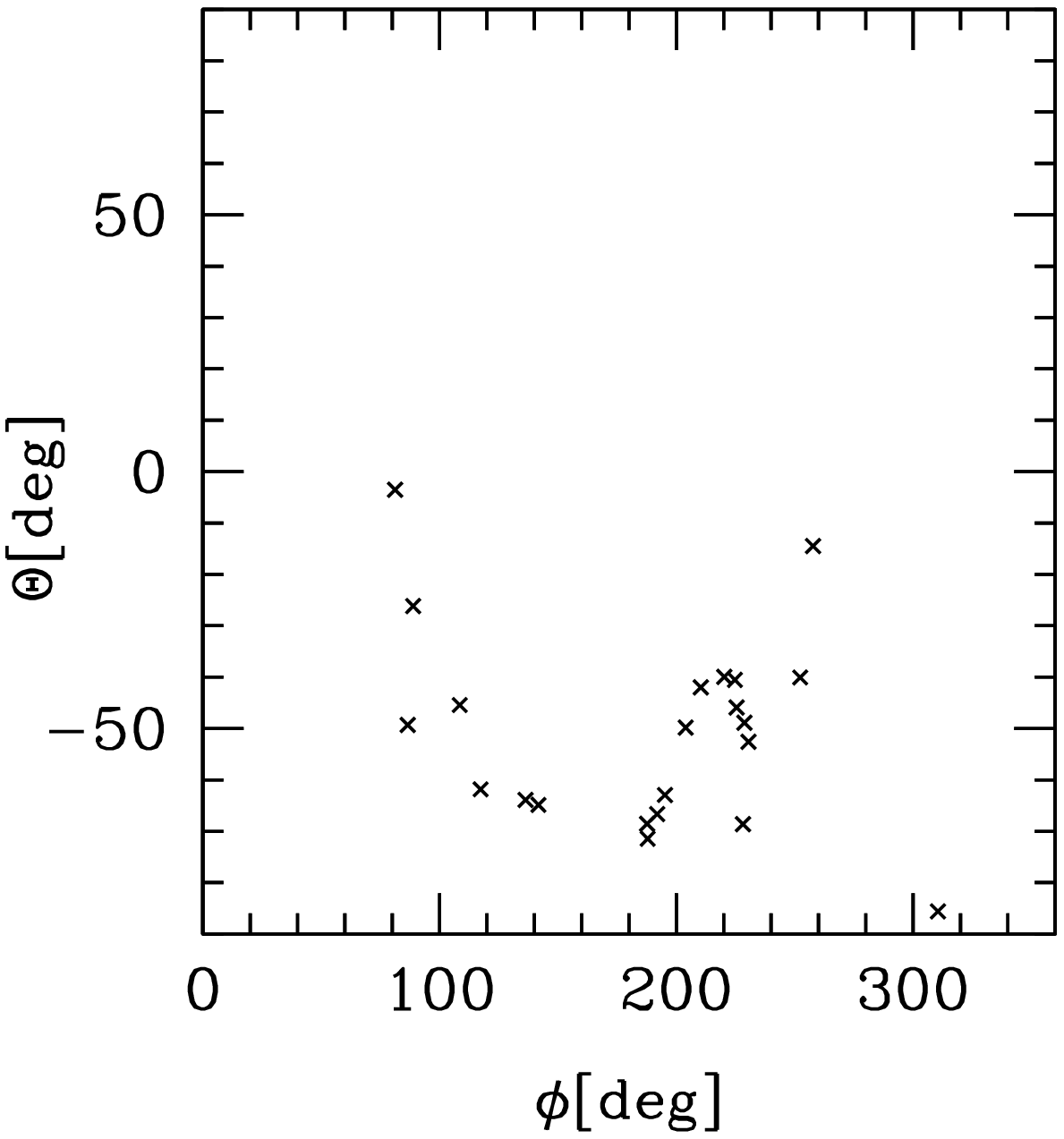}
    \epsfxsize=7.2cm
    \epsfbox[-20 0 340 370]{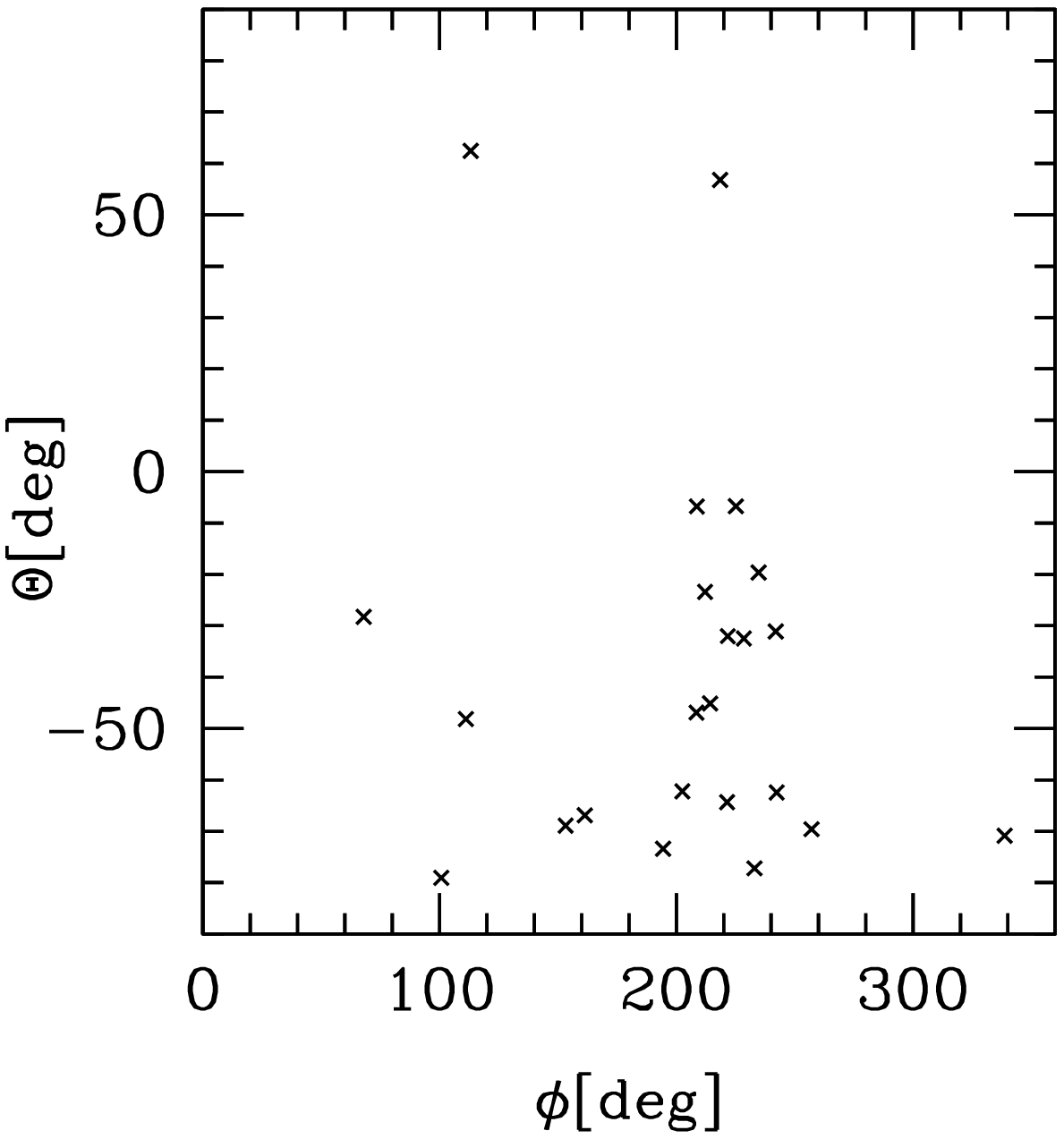}
\caption{The directions of the angular momentum vectors of LGA subhaloes with respect to M31. The
crosses indicate the angular coordinates of the vectors (in the spherical coordinates system) at the
time of infall, $t=9$ Gyr (upper panel) and at the last simulation output $t=13.4$ Gyr (lower panel).
The coordinate system was chosen so that the angular momentum vector of M31 points along the $z$ axis and the
orientation of $x$ and $y$ (different in each panel) so that the data points appear centrally in the plots.}
\label{lgaang}
\end{figure}

We tested different sets of parameters of the algorithm. We were looking
for a set of parameters which would recover whole complete groups but would
also reject haloes that were members of the group for a short time. Results are
quite stable and changing the linking length and the period by less than 30
per cent affects the group membership very little. Finally, two different sets of parameters of the group finder were
used. The first run for M31 was made with $l=100$ $h^{-1}$ kpc, $p=1.5$ Gyr, $t_1=0.2$ Gyr,
$t_2=13.4$ Gyr, thus in this case we look for groups in almost the whole duration of the simulation. The linking length
of $100$ $h^{-1}$ kpc is sufficient for finding large groups of subhaloes.

As a result we get one large group of 116 haloes and about 40 small groups with up to 10 haloes. It turns out that the
large group is indeed a bound structure, while the smaller groups are just accidentally linked. The large group was the
first target of the analysis. The next step was to narrow down the criteria of the group finder. In the second run we
used $l=100$ $h^{-1}$ kpc, $p=1.5$ Gyr, $t_1=7$ Gyr, $t_2=10.5$ Gyr. The linking length remains the same, but the
period at which the linking is made was reduced to the time between 7 and 10.5 Gyr which roughly corresponds to the
infall time of the large group on to M31. This way we were able to study the large group itself at a time close to its
accretion and later. The second run thus can be treated as a filter applied on the first run allowing to reject
haloes which deviated from the group before infall. At this point the results are even less influenced by the exact
values of the parameters.

\begin{figure*}
    \leavevmode
    \epsfxsize=15cm
    \epsfbox{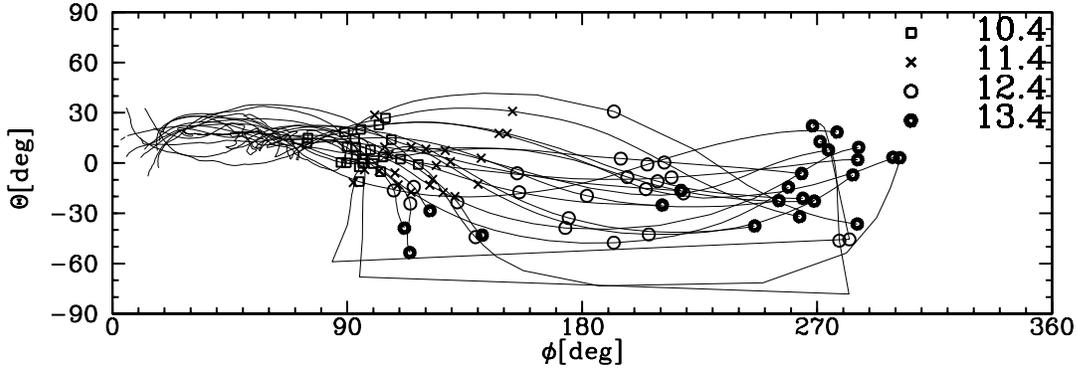}
\caption{Orbits of the LGA subhaloes as seen by the observer placed at the centre of M31. The spherical coordinate
system was oriented so that the angular momentum vector of the host points along the $z$ axis and the
orientation of the $x$ and $y$ axis was chosen so that the trajectories are continuous between $\phi=0$ and
$360^{\circ}$. Different symbols correspond to snapshots at different times (in Gyr) indicated in the legend.}
\label{lgasky}
\end{figure*}

\subsubsection{The large group of M31}

The upper panel of Fig.~\ref{histgroupsize} shows the histogram of the distribution of the group size following from
the second run of the group finding algorithm for M31. With the adopted parameters the largest group consists of
30 haloes. From now on we call it the Large Group of Andromeda (LGA). Few tens of smaller groups are also found. The
cumulative mass function of the LGA shown with the solid line in Fig.~\ref{mfgroups} illustrates its nature. The group
consists of one very large halo of mass $M=7.5 \times 10^{10}$ $h^{-1}$ M$_\odot$, well visible as the largest
subhalo in the image of M31 shown in Fig.~\ref{M31}, and several tens of smaller haloes which are actually its
satellites. Figure~\ref{lgapic} shows the spatial distribution of the haloes in the group at two different times. We
did not find any mergers inside such a group.

\subsubsection{Observational effects of a LGA-like group}

We now consider the present effects of an infall of a LGA-like group in the past. In this case we have a
large halo which falls on to M31 along with the set of its satellites. In the simulation the large
halo survives until the end, but eventually it will merge with M31. We are more interested in
the properties of the satellites. The question is whether we can detect such a past merger by studying only
the properties of dwarfs at present.

Figure~\ref{lgaang} shows the distribution of the angular momenta vectors of the LGA subhaloes with respect
to M31. Only the direction of the vector is shown in the spherical coordinate system. The
orientation of the system was chosen so that the $z$ axis points along the total angular momentum of M31. One can
clearly see that during the infall (upper panel) angular momenta of the subhaloes are correlated as postulated by Li \&
Helmi (2008) who defined groups by the separation of angular momenta vectors. In our case however only some of the
haloes show this pattern and several others have their angular momentum vectors differing more than the value of only
$10^\circ$ used by Li \& Helmi. This suggests that the method of group finding by comparing only angular momentum
orientations may miss larger objects or other bound subhaloes. The lower panel of Fig.~\ref{lgaang} shows the
distribution in the last simulation output, i.e. 4.4 Gyr later. This time period roughly corresponds to half of the
group orbit around M31, as can be seen in Fig.~\ref{lgapic}. During that time the angular momenta were modified and no
clear pattern seems to be preserved.

Figure~\ref{lgasky} shows the LGA haloes orbits projected on to the sky. The observer
was placed in the M31 centre and the coordinate system was again chosen so that the $z$ axis is aligned with the
angular momentum of M31. The haloes are infalling from one well-defined region in the sky but the
spread increases with time. It reaches a maximum around 12.4 Gyr when the haloes occupy almost half of the
sky. Then the haloes from the main group seem to fall back on the group again, but some other are left in the
opposite part of the sky. This example suggests that the alignment of angular momenta of infalling satellites
is not well conserved, even though the group has not yet decayed (see also Libeskind et al. 2007). One should
thus be very careful when trying to reproduce the histories of dwarf galaxies using their present proper
motions. Metz et al. (2007) studied the so-called disk of satellites in the LG. They found that the
distribution of MW and M31 satellites is not isotropic, and it rather forms a disk. It has been claimed that
this kind of structure could be an effect of a group infall on to the MW halo (Li \& Helmi 2008). Based on
our result we conclude that such a disk is probably not an effect of a group infall unless it happened very
recently.

\subsubsection{Smaller groups of M31}

The upper panel of Fig.~\ref{histgroupsize} shows that apart from the large group there is a number of smaller groups
consisting of usually two, sometimes several haloes. It is interesting to check whether haloes in such groups
could strongly interact with each other. We have calculated the escape velocities from such groups and compared them to
the true velocities of the haloes. It turns out that for the majority of the groups the subhaloes are only by chance
falling on to the host halo from the same direction and are not bound to each other. Several of such groups of two
haloes are marginally bound for very short periods of time. We studied in detail mass histories of such cases and
concluded that no significant mass transfer is present. These haloes are not massive enough to influence their
neighbourhood by tidal interactions and to stay bound for a longer period of time.

\subsection{Results of the group finding for MW}

The same algorithm was applied to the MW subhaloes. The analysis shows that there are no groups similar to
the LGA. The lower panel of Fig.~\ref{histgroupsize} shows the distribution of group sizes obtained with
$l=100$ $h^{-1}$ kpc, $p=1.5$ Gyr, $t_1=0.2$ Gyr, $t_2=13.4$ Gyr. We get a few tens of smaller groups and a larger one
consisting of 23 subhaloes, which we will call the Large Group of MW or LGM (with these parameters LGA had 116
members). A closer inspection shows that this group, although of different nature than LGA, is an interesting object by
itself. Figure~\ref{mfgroups} presents the cumulative mass function of this group with the dashed line. The group
consists of four larger subhaloes with masses around $M=10^9 h^{-1}$ M$_\odot$ and some smaller ones. Thus, in
contrast to LGA, this is a group of satellites of comparable size, rather than a single large halo with its subhaloes.
Unfortunately the infall time of the group is very late, it starts at around 12 Gyr from the beginning of the
simulation and is not completed by the end. This prevents us from studying the future evolution of such a group inside
the MW halo.

The group was formed around 4 Gyr from the beginning of the simulation. During most of the time it consists
of 12 haloes, while the rest is accreted late, when the group falls on to the MW. A careful analysis shows
that there is no significant mass transfer between large members. The smaller subhaloes tend to lose mass but
this process is much slower than for MW subhaloes and in total it does not amount to more than a few
percent of the initial mass. The larger subhaloes tend to gain mass both from smaller satellites and from
other infalling objects, but this process again is very slow and insignificant. We conclude that the
group members do not undergo any significant evolution inside the group.

\section{Mergers and interactions of the haloes}

In the previous section we have shown that being a member of a group of subhaloes does not significantly
affect the evolution of a given subhalo and thus cannot by itself lead to the formation of an early type
galaxy. Another possible channel by which such objects could form are interactions and direct mergers of subhaloes.
This issues have already been addressed to some extent by Knebe, Gill \& Gibson (2004) and Knebe et al. (2006) where
they found that on average 30 per cent of the substructure population experienced encounters and that such interactions
can account for a significant fraction of mass loss. In this section we address the question of how often such mergers
and interactions occur and whether they can indeed lead to the formation of dwarf spheroidals and ellipticals.

\subsection{Algorithm for finding interacting haloes}

The algorithm presented here was based on finding and selecting interacting haloes. We look for haloes between which
particles are exchanged. For each halo an interaction is defined by the following pair of parameters: the mass of all
the other haloes taking part in the encounter expressed as a fraction of the studied halo's mass, $m_1$, and the mass
fraction gained by the studied halo during the encounter, $m_2$. For example, the values $m_1=0.5, m_2=0.3$ mean that a
minimal interaction would be with haloes of mass equal to half the mass of the studied halo from which it gains 30
per cent. For an interaction to be interesting, $m_1$ has to be large enough, while $m_2$ needs to be some significant
fraction of mass of the other haloes. It is worth noting that a halo might become a subhalo of a larger halo, but then
leave it again (as found in cosmological simulations by e.g. Gill et al. 2005; Warnick, Knebe \& Power 2008; Ludlow et
al. 2009). In such a case the algorithm is capable of finding the interaction for both haloes even though it is not a
classical merger. The results will depend on the assumed parameters and the amount of matter exchanged.

\begin{figure}
    \leavevmode
    \epsfxsize=7.2cm
    \epsfbox[0 0 370 370]{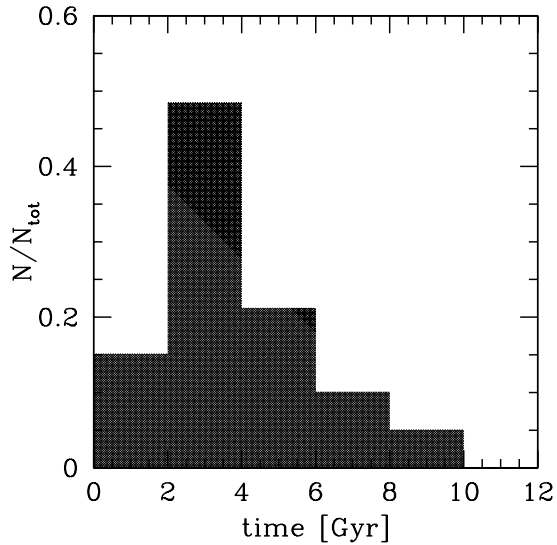}
    \caption{The distribution of interactions for the MW and M31 surviving subhaloes in time for the parameters
$m_1=0.1, m_2=0.3$. The number of interactions was expressed as a fraction of all interactions and the histogram was
normalized to unity.}
\label{histmerge}
\end{figure}

The motivation of this approach is the following. In most studies only major mergers are considered. However here we
are interested in the probability for a halo to potentially form a dwarf spheroidal or elliptical galaxy by an
interaction. Usually, a strong interaction between two haloes would eventually lead to a merger anyway. But the merger
itself could happen between haloes which have already exchanged large amounts of mass in past interactions. The
resulting merger could seem not to be significant. Our approach allows us to detect each of those interactions and
find the strongest one instead of studying only the merger itself. We would like to know what was the
magnitude of the strongest of these interactions, as this one had the largest chance to transform the galaxy.

\subsection{Interactions of the infalling haloes}

We consider only the interactions occurring for those haloes
which are still on their way to the host and have not yet become satellites. In
this analysis we include only those haloes of the MW and M31 which survived until the end of the
simulation with 100 particles or more, as only those are massive enough to be interesting from the observational point
of view. We check all detected events so a given surviving subhalo could have had more that one interaction in the past.
We also require that at the moment of interaction the studied halo needs to have at least 50 particles to avoid
numerical noise.

\begin{figure}
    \leavevmode
    \epsfxsize=7.2cm
    \epsfbox[0 10 340 370]{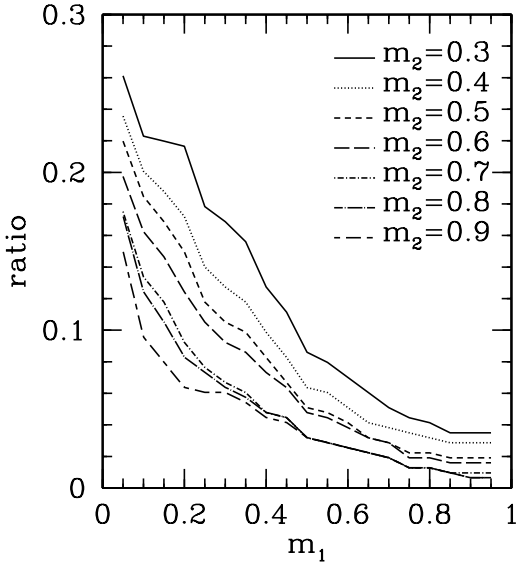}
    \caption{The dependence of the number of interactions on parameters $m_1$ and $m_2$. The lines
show the number of events that happened to the surviving sample of 198 M31 and 136 MW subhaloes as a
function of $m_1$. The different line types correspond to different values of $m_2$ as indicated by the legend. The
numbers give the ratio of the number of interactions to the total number of subhaloes.}
\label{merg}
\end{figure}

As already mentioned, at the end of the simulation M31 has 198 and the MW 136 surviving subhaloes of mass $M>2.5
\times 10^7$ $h^{-1}$ M$_\odot$. For example, if we assume that the set of parameters $m_1=0.5, m_2=0.3$ would be
characteristic for a major interaction and apply it, then the algorithm detects a total number of 28 interactions
for both M31 and MW subhaloes. This means that less than 10 percent of surviving satellites had a major interaction in
the past. This is a significant number, but not high enough to explain the origin of all dwarf spheroidal or elliptical
satellites. Figure~\ref{histmerge} shows a combined distribution of both M31 and MW interactions in time. Most of the
mergers occurred quite early, around redshift $z=2$ corresponding to 3 Gyr since beginning of the
simulation. Such a collision would not have to produce an early type dwarf galaxy immediately, since it would
still have a lot of time to evolve in the tidal field of its host.

Note that the shape of the distribution shown in Fig.~\ref{histmerge} does not depend strongly on the assumed
parameters $m_1$ and $m_2$; the interactions still occur rather early if we vary those parameters. The total number
of interactions detected depends however on these parameters. Figure~\ref{merg} presents the results of
exploring the parameter space in terms of the fraction of subhaloes that experienced interaction (or a number of
interactions per subhalo) with a given $m_1$ and $m_2$. As expected, minor interactions are more numerous and the
fraction of affected subhaloes decreases strongly with growing $m_1$ and $m_2$.

It turns out that a significant fraction (15-25 percent, taking the smallest $m_1$ and the whole range of $m_2$ in
Fig.~\ref{merg}) of subhaloes had some kind of interaction in the past, but only several percent had a strong
interaction that could be called a major merger. This suggests that the mergers, while present in the histories of some
dwarf galaxies, could not explain the large numbers of spheroidals and ellipticals in the LG. Still, majority of dwarf
galaxies did not have even a slight interaction with another dwarf in the past.

However, this result depends on the
assumption that we count the interactions only when the halo has at least 50 particles, corresponding to the mass
of $M=1.25 \times 10^7 h^{-1}$ M$_{\odot}$, which we find to be the lowest limit at which this study still makes sense.
We have verified that the assumed threshold of 50 particles for the
algorithm is rather realistic: with increasing threshold the numbers of interactions decrease very slowly, while
lowering the threshold would produce many more interactions. This could be easily understood: according to
Fig.~\ref{histmerge} most of the encounters are expected to happen early when the haloes are in general less massive.
Interactions with low particle numbers may however be affected by numerical noise and therefore their statistics is
not reliable.

\section{Discussion}

We have analyzed a constrained simulation of the local Universe which reproduces the main properties of the LG. The
simulation outputs were used to study the population of subhaloes around the largest galaxies. Our attention was
focused on those signatures of the evolution that can shed some light on the possible scenarios of the formation of
early type dwarf galaxies, dwarf ellipticals and dwarf spheroidals, that we presently observe. Assuming that the
progenitors of these objects are baryonic disks embedded in dark matter haloes, some mechanisms are required that could
transform the disks into bulges. In the case of purely gravitational interactions, such an evolution can occur by three
channels: the tidal stirring in the gravitational field of the host galaxy, interaction of small haloes infalling
together in groups and mergers between prospective subhaloes.

A quantitative study of the effect of tidal forces on a single subhalo usually requires a different
simulation setup and much higher resolution (e.g. Klimentowski et al. 2007, 2009). In the present context the
tidal stripping manifests itself in the statistical properties of our subhalo population. We found that the
mass function of satellites evolves significantly between $z=1$ and $z=0$ so that the number of small
subhaloes increases and the number of larger ones decreases, as expected if subhaloes lose mass by tides. On
average, the mass of satellites at the end of the simulation is smaller than at the moment of their infall on
to the halo of their host. For most subhaloes the transition from the phase of mass accretion to the phase of
mass loss can be easily identified and corresponds to the moment of entry into the host halo. We find that
this maximum mass or entrance moment can occur at a wide range of redshifts but typically more massive
subhaloes enter the vicinity of their hosts earlier, around $z=1$ and before (see also Diemand et al. 2007). This
distribution in redshift suggests that those massive subhaloes could be the progenitors of present-day dwarf
spheroidals since they have still a lot of time to evolve. The less massive subhaloes entering later could still be
dwarf irregulars today on one of their first passages around the MW like the Magellanic Clouds (Besla et al. 2007).

We also studied the evolution of satellites in groups. Although many groups were identified both around
M31 and the MW, they are usually loosely bound and their member subhaloes do not strongly interact with
each other. Therefore we conclude that being a member of a group cannot result in any morphological
transformation of a dwarf galaxy. Membership in a group could however explain some particular distributions
of satellites around big galaxies, provided that the group infall occurred recently. Using a large group
identified in the vicinity of our simulated M31 we have demonstrated that the group forms a coherent
structure at infall but dissipates on a rather short timescale.

Mergers between subhaloes offer another possibility for the formation of early type dwarfs. We found
that around 10 percent of present-day subhaloes of M31 and the MW underwent a
major interaction with another halo in the past. Most of the
events took place early on, around $7-11$ Gyr ago, when the objects
have not yet become satellites of big galaxies. This suggests that dwarf ellipticals could be a product of
mergers rather than tidal evolution, as they tend to be more isolated objects found further from the MW than
dwarf spheroidals.

\section*{Acknowledgments}

The simulations used in this work were performed at the Leibniz Rechenzentrum Munich (LRZ) and at the Barcelona
Supercomputing Centre (BSC), partly funded by the DEISA Extreme Computing Project (DECI) SIMU-LU. JK and
E{\L} acknowledge support by the Polish Ministry of Science and Higher Education under grant NN203025333 as well as by
the Polish-German exchange program of Deutsche Forschungsgemeinschaft. AK is supported by the MICINN through the Ramon
y Cajal programme. SG acknowledges a Schonbrunn Fellowship at the Hebrew University Jerusalem. LAMV acknowledges
financial support from Comunidad de Madrid through a PhD fellowship. GY is supported by the Spanish Ministry
of Education through research grants FPA2006-01105 and AYA2006-15492-C03. YH has been partially supported by the ISF
(13/08). We are grateful for the hospitality of the Astrophysikalisches Institut Potsdam where part of this work was
done. We acknowledge the LEA Astro-PF collaboration and the ASTROSIM network of the European Science Foundation
(Science Meeting 2387) for the financial support of the workshop `The local Universe: from dwarf galaxies to galaxy
clusters' held in Jab{\l}onna near Warsaw in June/July 2009, where this work was completed.

\end{document}